\documentclass[a4paper,fleqn,usenatbib]{mnras}

\usepackage{newtxtext,newtxmath}

\usepackage[T1]{fontenc}
\usepackage{ae,aecompl}
\usepackage[utf8]{inputenc}

\usepackage{graphicx}	% Including figure files
\usepackage{amsmath}	% Advanced maths commands
\usepackage{amssymb}	% Extra maths symbols
\usepackage{float}
\usepackage{tikz}

\newcommand{\GalA}{J0423A}
\newcommand{\GalB}{J0423B}
\newcommand{\GalC}{J0423C}
\newcommand{\GalD}{J0423D}

\title[ALMACAL III: Neutral, Molecular, and Ionised Gas around an Absorption-Selected System]{ALMACAL III: A combined ALMA and MUSE Survey for Neutral, Molecular, and Ionised Gas in an HI-Absorption-\\Selected System}

\author[A. Klitsch et al.]{A. Klitsch,$^{1,2}$\thanks{E-mail: aklitsch@eso.org (A. Klitsch)}
C. P\'eroux,$^{3}$ 
M. A. Zwaan,$^{1}$ 
I. Smail,$^{2}$ 
I. Oteo,$^{1}$ 
A. D. Biggs,$^{1}$ 
\newauthor{G. Popping$^{1,4}$ 
and A. M. Swinbank$^{2}$}\\
$^{1}$European Southern Observatory, Karl-Schwarzschild-Str. 2, 85748 Garching near Munich, Germany\\
$^{2}$Centre for Extragalactic Astronomy, Durham University, Department of Physics, South Road, Durham DH1 3LE, UK\\
$^{3}$Aix Marseille Univ, CNRS, LAM, (Laboratoire d'Astrophysique de Marseille), UMR 7326, 13388, Marseille, France\\
$^4$Max-Planck-Institut für Astronomie, Königstuhl 17, 69117 Heidelberg, Germany}

\date{Accepted XXX. Received YYY; in original form ZZZ}

\pubyear{2017}

\begin{document}
\label{firstpage}
\pagerange{\pageref{firstpage}--\pageref{lastpage}}
\maketitle

\begin{abstract}

Studying the flow of baryons into and out of galaxies is an important part of understanding the evolution of galaxies over time. We present a detailed case study of the environment around an intervening Ly~$\alpha$ absorption line system at $z_{\rm abs} = 0.633$, seen towards the quasar J0423$-$0130 ($z_{\rm QSO} = 0.915$). We detect with ALMA the \mbox{$^{12}$CO(2--1)}, \mbox{$^{12}$CO(3--2)}  and $1.2$~mm continuum emission from a galaxy at the redshift of the Ly~$\alpha$ absorber at a projected distance of $135$~kpc. From the ALMA detections, we infer ISM conditions similar to those in low redshift Luminous Infrared Galaxies.
DDT MUSE integral field unit observations reveal the optical counterpart of the $^{12}$CO emission line source and three additional emission line galaxies at the absorber redshift, which together form a galaxy group. The $^{12}$CO emission line detections originate from the most massive galaxy in this group. While we cannot exclude that we miss a fainter host, we reach a dust-uncorrected star-formation rate (SFR) limit of > $0.3~\text{M}_{\sun} \text{ yr}^{-1}$ within $100$~kpc from the sightline to the background quasar. We measure the dust-corrected SFR (ranging from $3$ to $50$~M$_{\sun}$~yr$^{-1}$), the morpho-kinematics and the metallicities of the four group galaxies to understand the relation between the group and the neutral gas probed in absorption. We find that the Ly~$\alpha$ absorber traces either an outflow from the most massive galaxy or intra-group gas. This case study illustrates the power of combining ALMA and MUSE to obtain a census of the cool baryons in a bounded structure at intermediate redshift.

\end{abstract}


%%%%%%%%%%%%%%%%%%%%%%%%%%%%%%%%%%%%%%%%%%%%%%%%%%

%%%%%%%%%%%%%%%%% BODY OF PAPER %%%%%%%%%%%%%%%%%%

\section{Introduction}

A key part of understanding the evolution of galaxies is to determine how gas is accreted onto galaxies and how it is exchanged with their surroundings via outflows. Since ultimately the respective evolutions of gas, stars, and metals are connected through the stellar life cycle, gas flows have an immediate impact on the history of star formation~(SF) and the chemical enrichment. Inflows of pristine gas might dilute the metal content of galaxies and certainly fuel their SF, whereas outflows take away the metals, chemically enriching the circum-galactic medium (CGM). High column density gas lying along a quasar line of sight is an ideal tool to study the CGM. These gas clouds characterized by Ly~$\alpha$ or metal absorption profiles can probe not only a static CGM, but also flows of gas through the halo. To obtain a full picture of the CGM and its interplay with galaxies, it is important to study the location, kinematics, and metallicity of both the absorbing gas and the nearby galaxy. 

Using integral field spectroscopy (IFS) it is possible to efficiently identify galaxies at the redshift of absorbers as well as determine their star-formation rate (SFR), kinematics and metallicity \citep[e.g.][]{Bouche2007sinfoni, Peroux2011sinfoni, Jorgenson2014spatially, Peroux2017nature}. Interestingly, recent studies have shown that the identification of the Ly~$\alpha$ absorber host galaxy is not always unique. In some cases, the absorber can be linked to one galaxy, but in other cases, it is associated with intra-group gas \citep{Whiting2006mos, Kacprzak2010galaxy, Gauthier2013ultra, Rahmani2017observational, Bielby2017probing, Fumagalli2017witnessing, Peroux2017nature}.

Recent studies have shown that the CGM extends to at least $100$~kpc for isolated galaxies \mbox{\citep{Prochaska2017cos}} and $140$~kpc in groupdf of galaxies \mbox{\citep{Bordoloi2011radial}}. These distances convert to a radial angular extent of $\sim20\arcsec$ at intermediate redshifts of $z = 0.5$ requiring a large field of view integral field unit (IFU) such as MUSE to cover the full potential extent of the CGM.

Large surveys aiming to understand the connection between absorption line systems and their host galaxies either use galaxy-quasar pairs searching for absorption from the known galaxy in the quasar spectrum \citep[e.g.][]{Tumlinson2013cos} or target known absorbers and search for the associated galaxies using integral field spectroscopy \citep[e.g.][]{Schroetter2016Megaflow, Rahmani2017observational, Bielby2017probing, Fumagalli2017witnessing, Peroux2017nature}. Outflows are ubiquitously observed at all redshifts \citep[e.g.][]{Rupke2005outflows, Veilleux2005galactic, Tremonti2007discovery, Weiner2009ubiquitous} %(e.g. Pettini et al. 2001; Steidel, Pettini & Adelberger 2001; Martin 2005; Rupke, Veilleux & Sanders 2005; Veilleux, Cecil & Bland-Hawthorn 2005; Tremonti et al. 2007; Weiner et al. 2009) 
while direct observations of inflows are less commonly found \citep[e.g.][]{Martin2012demographics, Bouche2013Signatures}, %Sato et al. 2009; Martin et al. 2012; Bouch´e et al. 2013; Diamond-Stanic et al. 2015
possibly because they are more difficult to observe.

A completely new perspective is opened by studying the molecular gas of the Ly~$\alpha$ absorber host galaxy traced by $^{12}$CO emission lines. Recently, \mbox{\citet{Neeleman2016first}} reported the first detection of $^{12}$CO(1-0) emission from such a host galaxy at $z=0.101$. Moreover, the combination of $^{12}$CO emission line detections with IFS, potentially allows us to get a complete census of the stars and the cool gas in such systems. This is the next important step towards a better understanding of such systems and therefore the flow of baryons through galaxies. %Moreover, the detection of multiple $^{12}$CO emission lines enables us to study the physical properties of the star -forming gas.

In our new (sub)mm survey ALMACAL, we detect for the first time multiple $^{12}$CO transitions from a galaxy first identified as an intervening absorber at $z_{\rm abs} = 0.633$ towards the quasar J0423$-$0130 ($z_{\rm QSO} = 0.915$). We have obtained additional MUSE observations revealing a group of four galaxies at the absorber redshift, where one of these is coincident with the $^{12}$CO emission seen in the ALMA observations. The immediate aim of this study is to identify the origin of the absorption seen towards the quasar. In a broader perspective, this system serves us as a reference system to demonstrate the need for a multi-wavelength study of intervening absorbers and their environment.

This paper is organized as follows: In \S~2 we describe our dataset from the ALMA archive, the new MUSE observations, the ancillary archival data and previous work. We describe the source detections and measurements from the ALMA and MUSE observations, the broad-band photometry based on archival data, and the analysis of the measurements in \S~3. In \S~4 we discuss possible scenarios explaining the absorption. A summary and the conclusions are given in \S~5. We present additionally morpho-kinematic modelling and the fitting of the absorption lines in the quasar spectrum in the Appendix. Throughout the paper, we adopt an H$_0=70$~km~s$^{-1}$~Mpc$^{-1}$, $\Omega_{\rm M}=0.3$, and $\Omega_{\Lambda}=0.7$ cosmology.

%\section{Archival Data and New Observations of the Field of J0423$-$013}
\section{Observations and Data Reduction}

\begin{figure*}
\begin{tikzpicture}
    \node[anchor=south west,inner sep=0] (image) at (0,0) {\includegraphics[width = \linewidth, trim = 330 10 0 0, clip]
{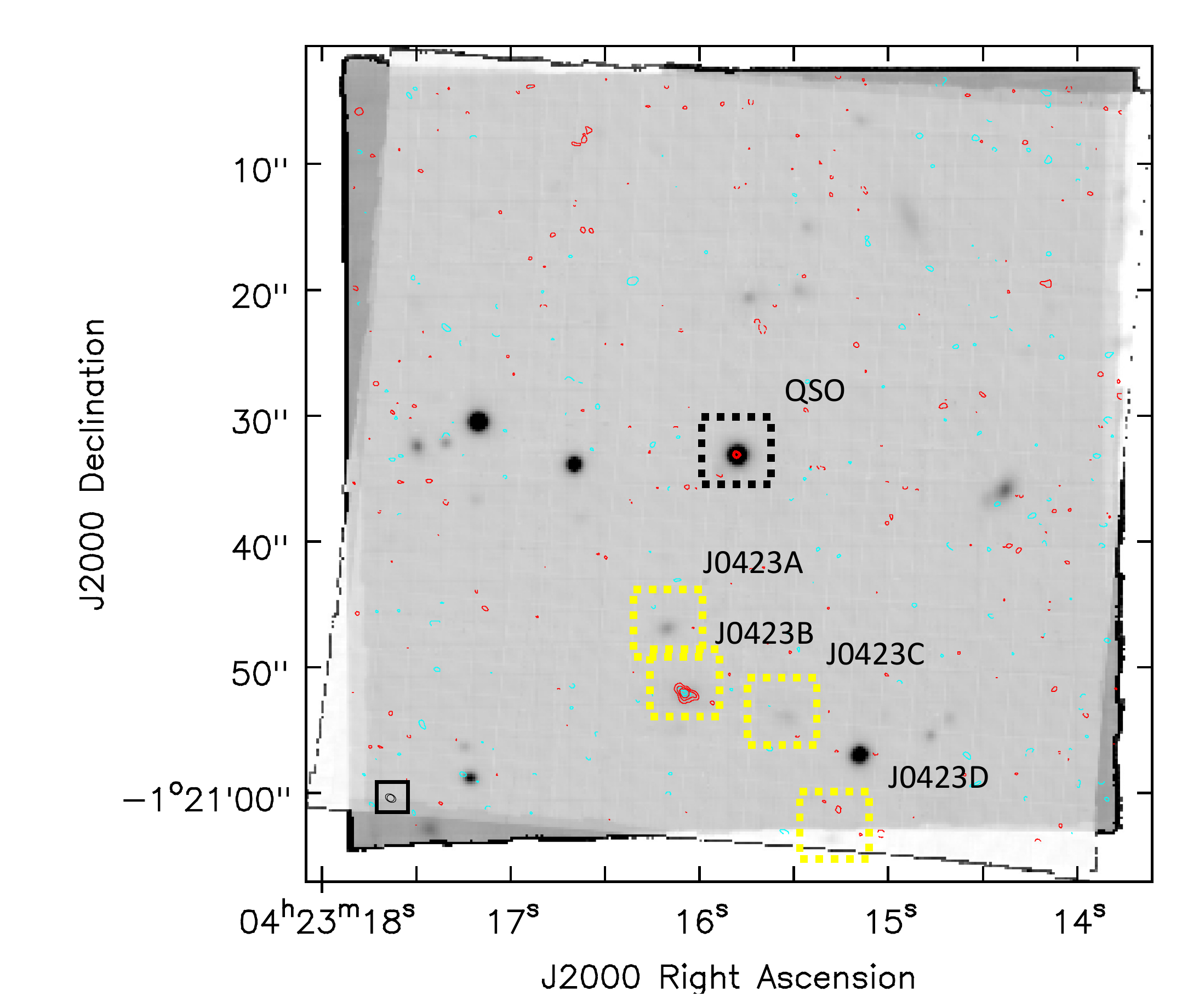}};
    \begin{scope}[x={(image.south east)},y={(image.north west)}]
       % \draw [ultra thick] (0.1,0.1) -- node[below]{\Large 50 kpc} (0.21,0.1);
        \draw[dashed] (0.61,0.55) circle (4.4cm);
		\draw[dashed] (0.61,0.55) circle (2.4cm);
    \end{scope}
\end{tikzpicture}
\caption{Overview of the detected sources in the field of J0423$-$0130. The reconstructed white-light image from our MUSE observations is shown in grey-scales, the red and cyan contours show the $^{12}$CO(2--1) and $^{12}$CO(3--2) emission, respectively. The contours show the $-3$, $3$, $5$, and $7\sigma$ levels in the respective mapdf, where negative contours are dashed. The galaxies identified to be at the absorber redshift based on out MUSE and ALMA observations are marked with the yellow boxes, the quasar is marked with the black box. The galaxies $1$ and $2$ in Table~\ref{TabCoordPrevStud} are not detected in our observations and therefore not shown in the figure and not considered in the analysis presented here. The dashed large and small circles mark the ALMA half power beam width in Band~4 and Band~6, respectively. The small black circles in the bottom left corner show the synthesized beams of the ALMA Band~4 and Band~6 observations which have a comparable size. A zoom in on J0423B is shown in Fig.~\ref{FigMUSEALMA}. The second OB of the MUSE observations was rotated by $5\degr$ with respect to the first OB due to a technical problem.}
\label{Fig.Overview1}
\end{figure*}

The aim of this work is to identify and characterize galaxies connected to the intervening quasar absorber with a particular focus on the molecular gas content of the host galaxies. 
To do this, we use ALMACAL, a novel wide and deep (sub-) millimetre survey utilizing calibrator observations from the ALMA archive \mbox{\citep{Oteo2016ALMACAL}}.

From our parent sample we have identified a particular absorber at $z = 0.633$ towards the quasar J0423$-$0130 ($04$h$23$m$15.8$s $-01$d$20$m$33$s, $z = 0.915$). The absorber is a Lyman Limit System with an HI column density of $\log (N(\rm{HI}) /\: \text{atoms} \: \text{cm}^{-2}) = 18.54^{+0.07}_{-0.10}$. We choose this source for our pilot study, because we detect two $^{12}$CO emission lines in our ALMACAL data. We have obtained MUSE follow-up observations of the field. We show that combining these two datasets with other ancillary data yields a more complete picture of such systems.

\subsection{ALMACAL}
\label{SecALMACAL}

\begin{table}
\caption{Summary of the used ALMACAL observations of J0423$-$0130}
\label{TabObsPropALMA}
\begin{minipage}{\linewidth}
\begin{tabular}{l l l l l l l }
\hline
Project code & Date & Scan & Band & $^{12}$CO & t$_{\rm exp}$\\
& & Intent& & Trans. & [s]\\
\hline
2015.1.00920.S & 09.01.2016 & B & 4 & 2$-$1 & $604.8$ \\
2015.1.00262.S & 22.08.2016 & B & 4 & 2$-$1 & $728.5$ \\
2012.1.00350.S & 04.12.2013 & B & 6 & 3$-$2 & $302.4$\\
2013.1.00403.S & 13.06.2014 & B & 6 & 3$-$2 & $302.4$\\
2013.1.00111.S & 05.07.2015 & B & 6 & - & $302.4$\\
2013.1.01175.S & 19.07.2015 & B & 6 & - & $605.8$\\
2013.1.01175.S & 19.07.2015 & F & 6 & - &  $302.4$\\
2013.1.01225.S & 05.08.2015 & B & 6 & - & $302.4$\\
2012.1.00146.S & 28.05.2015 & F & 6 & - & $151.2$\\
2015.1.00350.S & 27.10.2015 & B & 6 & - & $302.4$\\
2013.1.00198.S & 06.06.2015 & B & 6 & - & $302.4$\\
2013.1.01172.S & 30.06.2014 & B & 6 & - & $483.8$\\
2013.1.00815.S & 28.06.2014 & F & 6 & - & $151.2$\\
2013.1.00815.S & 16.06.2014 & B & 6 & - & $604.8$\\
2013.1.00710.S & 12.12.2014 & F & 6 & - & $151.2$\\
2015.1.00920.S & 01.01.2016 & B & 6 & - & $302.4$\\
2016.1.00724.S & 27.12.2016 & F & 6 & - & $151.2$\\
2016.1.00683.S & 29.11.2016 & F & 6 & - & $151.2$\\
2016.1.00627.S & 03.12.2016 & F & 6 & - & $151.2$\\
2016.1.00627.S & 01.12.2016 & F & 6 & - & $302.4$\\
2016.1.01453.S & 22.11.2016 & F & 6 & - & $151.2$\\
2016.1.01262.S & 30.11.2016 & F & 6 & - & $151.2$\\
2015.1.00296.S & 22.06.2016 & F & 6 & - & $151.2$\\
\hline
\end{tabular}
\end{minipage}
\begin{minipage}{\linewidth}
Note: Scan intent B denotes the bandpass calibrator, and F denotes the flux calibrator.
\end{minipage}
\end{table}

\begin{table*}
\caption{Summary of the final J0423$-$0130 ALMA data cube properties.}
\label{TabImPropALMACAL}
\begin{minipage}{\linewidth}
\centering
\begin{tabular}{l l l l l l}
\hline
Band & Freq & med. Ang. Res. & r.m.s. & $\Delta v$ & PB\\
 & [GHz] & [\arcsec] & [mJy~beam$^{-1}$] & [km s$^{-1}$] & FWHM [\arcsec]\\
\hline
4 & $141.166$ & $0.43$ & $0.28$ & $66$ & $48$\\
6 & $211.742$ & $0.70$ & $0.37$ & $44$ & $27$\\
\hline
\end{tabular}
\end{minipage}
\end{table*}

The data used in this work are taken from the ALMACAL survey, a wide and deep survey utilizing the ALMA calibrator archive. The retrieval and calibration of the ALMA calibrator data are described in detail by \mbox{\citet{Oteo2016ALMACAL}}.

Up until 1. July 2017, the quasar J0423$-$0130 was observed four times at frequencies corresponding to possible $^{12}$CO emission lines at the absorber redshift.
Two observations are available in Band~4 covering the frequency of the redshifted \mbox{$^{12}$CO(2--1)} emission line and two observations in Band~6 covering the frequency of the redshifted \mbox{$^{12}$CO(3--2)} emission line. The details of these observations are given in Table \ref{TabObsPropALMA}. The total integration time is $1333$~s in Band~4 and $605$~s in Band~6 for the observations covering the $^{12}$CO emission lines. Furthermore, we use all well calibrated Band~6 observations that do not cover the $^{12}$CO(3--2) emission line to search for continuum emission. The corresponding datasets are listed in Table~\ref{TabObsPropALMA}.

We carry out the calibration and data reduction using the Common Astronomy Software Applications ({\sc CASA}) software package version 4.7. We examine each individual dataset used for the emission line search first in the $uv$ plane to apply further interactive flagging if necessary. We reduce the spectral resolution to a maximum of $15.625$~kHz. This results in a velocity resolution of $33$~km~s$^{-1}$ (and $66$~km~s$^{-1}$ after Hanning smoothing) for the Band~4 observations and a velocity resolution of $22$~km~s$^{-1}$ (and $44$~km~s$^{-1}$ after Hanning smoothing) for the Band~6 observations. The spectral resolution ensures that an emission line with a width of $\gtrsim300$~km~s$^{-1}$ is covered by at least $5$ and $7$ channels in the respective bands.
Before the imaging, we concatenate the two datasets in each band.

For the image cubes and the continuum image, the imaging was done using the standard `clean' algorithm. A `robust' weighting scheme was applied, using a Briggs parameter of $0.5$, which guarantees a nearly optimal sensitivity while still providing a high spatial resolution and a well-behaved synthesized beam. In Band~4 we have two observations with a spatial resolution of $1.84\arcsec$ and $0.39\arcsec$ and therefore we use an outer taper of $0.5\arcsec$ to prevent too high a weighting of the higher resolution data. The final parameters of the image cubes are given in Table~\ref{TabImPropALMACAL}. The resulting mean r.m.s. noise level in Band~4 is measured to be $\sim0.28$~mJy~beam$^{-1}$ per $66$~km~s$^{-1}$ and the median angular resolution of the final data cube is $\sim0.43\arcsec$. The r.m.s. noise level in the Band~6 image cube is $\sim0.37$~mJy~beam$^{-1}$ per $44$~km~s$^{-1}$ and the median angular resolution of the final data cube is $\sim0.70\arcsec$. The r.m.s. noise level in the $1.2$~mm continuum image is $\sim96$~mJy~beam$^{-1}$~km~s$^{-1}$ and the median angular resolution is $\sim 0.75 \arcsec$. We use a pixel size of $0.15\arcsec$ in the Band~4 image cube, $0.24\arcsec$ in the Band~6 image cube covering the $^{12}$CO(3--2) emission line and $0.15\arcsec$ in the Band~6 continuum image cube.

\subsection{New MUSE Observations}

%\begin{table}
%\begin{minipage}{\linewidth}
%\caption{Journal of observations for the MUSE data}
%\label{TabJoO}
%\begin{tabular}{l l l l }
%\hline
%Project code & obs dates & int. time & seeing \\
%\hline
%ESO 298.A-5017(A) & 23.12.2016 & $2810$s & $0.56\arcsec$\\
% & 24.12.2016 & $2810$s & $0.71\arcsec$\\
% \hline
%\end{tabular}
%\end{minipage}
%\end{table}

We obtain Multi Unit Spectroscopic Explorer (MUSE) observations in the field of J0423$-$0130 through Director's Discretionary Time. Although the previously identified galaxies from our ALMACAL observations and another galaxy identified in the broad-band imaging by \mbox{\citet{Rao2011groundbased}} are located south of the quasar, we centre the field of view on the quasar in order to not exclude a priori the possibility to find a galaxy north of the quasar. The observations were carried out in service mode under programme ESO 298.A-5017 at the European Southern Observatory using MUSE at the Nasmyth focus of the $8.2$~m Very Large Telescope UT4. Two ``Observing Blocks'' (OBs) were taken on the 23 and 24 December 2016. The field is rotated by $180$~degrees between these two OBs. The OBs are further divided into two sub-exposures, with an additional rotation of $90$~degrees and a sub-arcsec dithering pattern. 
The exposure time of each sub-exposure is 1405s with a seeing of $0.55\arcsec$ - $0.7\arcsec$ resulting in a total exposure time of $5620$~s. The spectral coverage is $\sim4800$~--~$9300$~\AA. At the redshift of the absorber ($z_{\rm abs}=0.633$) we, therefore, cover all emission lines between [OII] and [OIII], but not H~$\alpha$. 

We reduce the data with v1.6.1 of the ESO MUSE pipeline and use the Zurich Atmosphere Purge~(ZAP) code for the sky subtraction \mbox{\citep{Soto2016Zap}}. % and use additional external software for source finding, to extract the 1D spectra and for the modelling of the kinematics.
We use the master bias, master flat field and arc lamp exposures taken closest in time to the science observations to correct the raw data cubes. This is done with the {\sc scibasic} recipe. We do the correction to a heliocentric reference system using the {\sc scipost} recipe. In this step, we do not perform a sky-removal since ZAP is best performing on non-sky-subtracted cubes. We check the wavelength solution using the position of the brightest sky lines and find it to be accurate to $10$~km~s$^{-1}$. The offset table is created for each cube by comparing the positions of point sources in the field using the recipe {\sc exp\_align}. In the final step, the cubes are combined using the recipe {\sc exp\_combine}. 
The seeing of the final cube measured from the quasar and other point sources is $0.71\arcsec$ at $7000$~\AA. To ensure a precise astrometry, we match the position of the quasar with its high precision position known from Very Long Baseline Array (VLBA) \mbox{\citep{Lanyi2010celestrial}}.

We remove the sky emission lines using the {\sc ZAP} code \mbox{\citep{Soto2016Zap}}. To determine regions with pure sky emission a mask is created from the reconstructed white light image. In this step, we set the flux levels accordingly to include bright objects in the mask. {\sc ZAP} isolates the emission lines caused by sky emission using a principle component analysis and removes them from the data cube. 

The quasar in this field is highly variable at optical wavelengths and we do not cover any other standard star with our field, so we estimate the flux error to be $\pm10\%$ based on our previous experience \citep{Rahmani2017observational, Peroux2017nature}.

\subsection{Ancillary Data}

\begin{table*}
\caption{Summary of all previously detected galaxies connected to the absorber towards J0423-0130.}
\label{TabCoordPrevStud}
\begin{tabular}{l l l l l l l}
\hline
ID & RA (J2000) & Dec (J2000) & $\theta$~[kpc] & Previous Detection & Ref. & Detected in this Work\\
\hline 
1 & n.a. & n.a. & 14.7 & broad-band & \citet{Churchill1996spatial} & No\\
2 & 04 23 15.58 & $-$01 20 34.6 & $24.7$ & broad-band & \citet{Rao2011groundbased} & No\\
3 & 04 23 16.18 & $-$01 20 46.9 & $102$ & [OII] & \citet{Yanny1992emission}, & Yes - \GalA\\ 
& & & & &\citet{Rao2011groundbased}\\
4 & 04 23 16.07 & $-$01 20 52.1 & $133$ & [OII] & \citet{Yanny1992emission} & Yes - \GalB\\
5 & 04 23 15.54 & $-$01 20 54.0 & $146$ & [OII] & \citet{Yanny1992emission} & Yes - \GalC\\
6 & 04 23 15.30 & $-$01 21 03.7 & $216$ & [OII] & \citet{Yanny1992emission} & Yes - \GalD\\
\hline
\end{tabular}
\begin{minipage}{\linewidth}
Note: For galaxy 1 \mbox{\citet{Churchill1996spatial}} only report the impact parameter, but not the coordinates. \mbox{\citet{Yanny1992emission}} do not give the coordinates of the detected galaxies. Therefore, we match the galaxies with our observations and quote the coordinates based on this.
\end{minipage}
\end{table*}

The target studied in this paper is the intervening metal line absorber initially detected by \mbox{\citet{Wills1980Observations}} as an FeII and MgII absorber at $z_{\rm abs} = 0.6320$ towards the quasar QSO J0423$-$0130 $(z_{\rm QSO} = 0.915)$.

\subsubsection{Early Observations}

This system was observed by several authors using narrow-band [OII] filters \mbox{\citep{Yanny1990emission, Yanny1992emission}} and broad-band imaging \mbox{\citep{Churchill1996spatial, Rao2011groundbased}} to determine the galaxy associated with the intervening absorber. In total six systems were reported with an impact parameter $\lesssim 200$~kpc. %, where we do not detect two of these objects. 
However, no consensus was reached on what is causing the absorption. The coordinates of these galaxies are listed in Table~\ref{TabCoordPrevStud}. 

\subsubsection{Quasar Spectroscopy}

The strong FeII and MgII absorber was reobserved by \mbox{\citet{Churchill1996spatial}} as part of a program to study the spatial and velocity distribution of absorbing systems with known galaxy counterparts. The authors obtained a high-resolution spectrum of J0423$-$0130 in January 1995 using the HIRES echelle spectrometer \mbox{\citep{Vogt1994HIRES}} on the Keck 10m telescope. The spectral resolution is $6.6$~km~s$^{-1}$. 

The absorber towards J0423$-$0130 was also part of an MgII-FeII absorber sample, which was followed up by \mbox{\citet{Rao2006damped}} using the  \textit{Hubble Space Telescope (HST) Space Telescope Imaging Spectrograph (STIS)} spectrograph in Cycle 9 (PID 8569). The HI column density was determined to be $\log( N(\rm{HI}) /\: \text{atoms} \: \text{cm}^{-2}) = 18.54^{+0.07}_{-0.10}$.

\subsubsection{Imaging of the Field of J0423-0130}

\mbox{\citet{Churchill1996spatial}} obtained broad-band imaging for the field of J0423$-$0130 and reported a galaxy associated with the absorber at an impact parameter of $14.7$~kpc.\\
Furthermore, \mbox{\citet{Rao2011groundbased}} observed this field using the MDM Observatory $2.4$~m Hiltner telescope. They analysed a $30\arcsec \times 30\arcsec$ wide region of their images corresponding to a field of view of $205 \times 205$~kpc at the absorber redshift. The images were taken in the \textit{B}, \textit{R}, \textit{I}, \textit{J}, \textit{H}, and \textit{K} bands. The pdfF of the quasar was only subtracted in the optical bands. \mbox{\citet{Rao2011groundbased}} performed a stellar population synthesis modelling for the five sources they detected within a $100$~kpc radius around the quasar. It was found that only one galaxy has a photometric redshift of $z = 0.637 \pm 0.031$ consistent with the absorber ($z_{\rm abs} = 0.6331$). The impact parameter of this galaxy is $14.5\arcsec$ or $99.6$~kpc. Furthermore, it was reported that another source might be at an impact parameter of $3.6\arcsec$, which would translate to $25$~kpc at the absorber redshift. However, it was only visible in the infrared data and the spatial resolution prevented the isolation of the object from the quasar in these frames. Therefore, \mbox{\citet{Rao2011groundbased}} could not perform stellar population synthesis modelling for this source. The coordinates of the detected galaxies are given in Table~\ref{TabCoordPrevStud}. In the analysis we use the broad-band images to obtain broad-band photometry of the MUSE detected galaxies.

\section{Analysis}

\subsection{ALMA Source Detection and Flux Measurement}
\label{SubsecALMAdata}

\begin{figure}
\includegraphics[width = \linewidth]{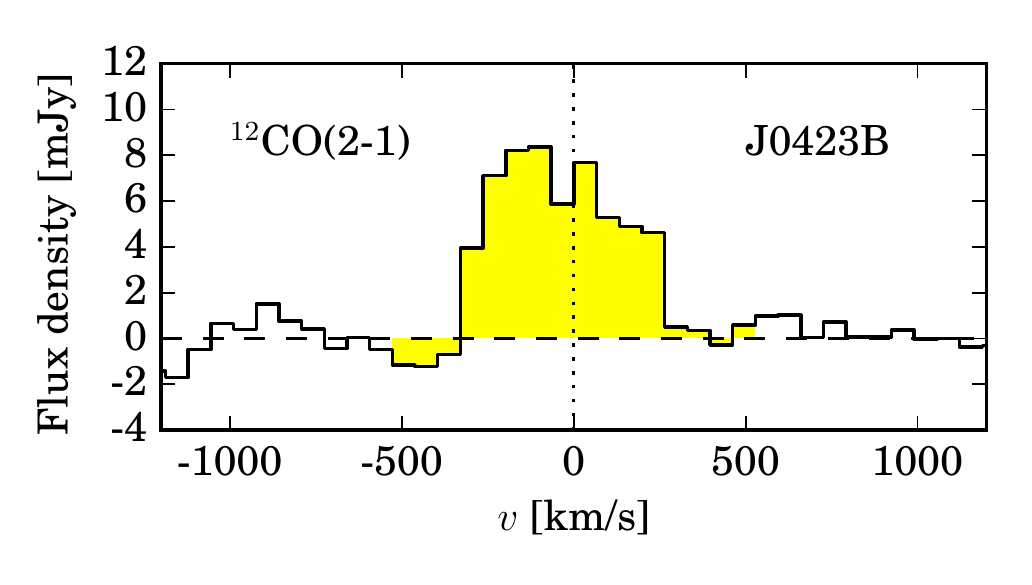}\vspace{-1.2cm} \\
\includegraphics[width = \linewidth]{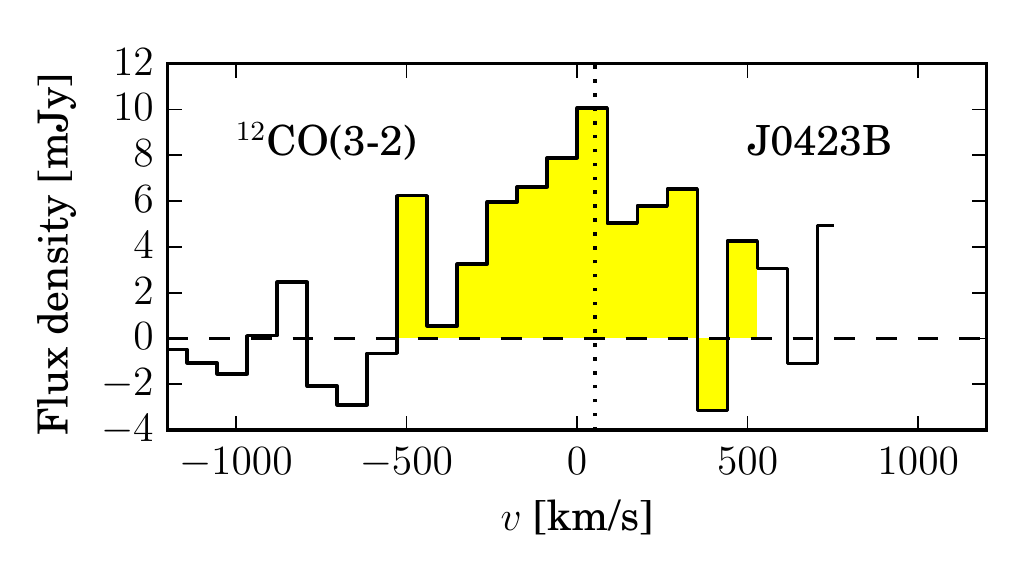}
\caption{$^{12}$CO(2--1) (top) and $^{12}$CO(3--2) (bottom) emission lines observed in our ALMACAL data cubes centred in position on the quasar J0423-0130. The zero velocity corresponds to the redshift determined from the respective $^{12}$CO emission line. The yellow region marks velocity range over which we integrate the total flux. The dotted lines show the position of the main absorption component at a redshift of $0.633174$. The peak in the $^{12}$CO emission line redshift corresponds exactly with that of the Lyman alpha absorption line, which signifies the clear association between the absorption system and the $^{12}$CO gas. The spectra are binned to a resolution of 66 and 88~km~s$^{-1}$ for the $^{12}$CO(2--1) and $^{12}$CO(3--2), respectively.}
\label{FigCOspectra}
\end{figure}

\begin{table*}
\caption{ALMA detection of the galaxy \GalB}
\begin{tabular}{llllll}
\hline
Line & $z$ & S$_{\rm peak}$ & S$_{\rm int}$& $L'_{\rm CO}$& $FWHM$\\
 & & [mJy] & [Jy km s$^{-1}$] & [$ \times10^{-10}$ K km s$^{-1}$ pc$^{2}$]&  [km s$^{-1}$]\\
\hline
$^{12}$CO(2--1) & $0.6331$7 & $8.4 \pm 0.7$ & $3.6 \pm 0.2$ & $1.9 \pm 0.1$ & $590 \pm 30$\\
$^{12}$CO(3--2) & $0.63335$ & $10 \pm 1$ & $5.2 \pm 0.3 $ & $1.2 \pm 0.1$ & $610 \pm 40$\\
\hline
\end{tabular}
\label{TabCOemissionSummary}
 \end{table*}

\begin{figure*}
\centering
\begin{tikzpicture}
    \node[anchor=south west,inner sep=0] (image) at (0,0) {\includegraphics[width = 0.48\linewidth, trim = 10 400 10 10 crop]{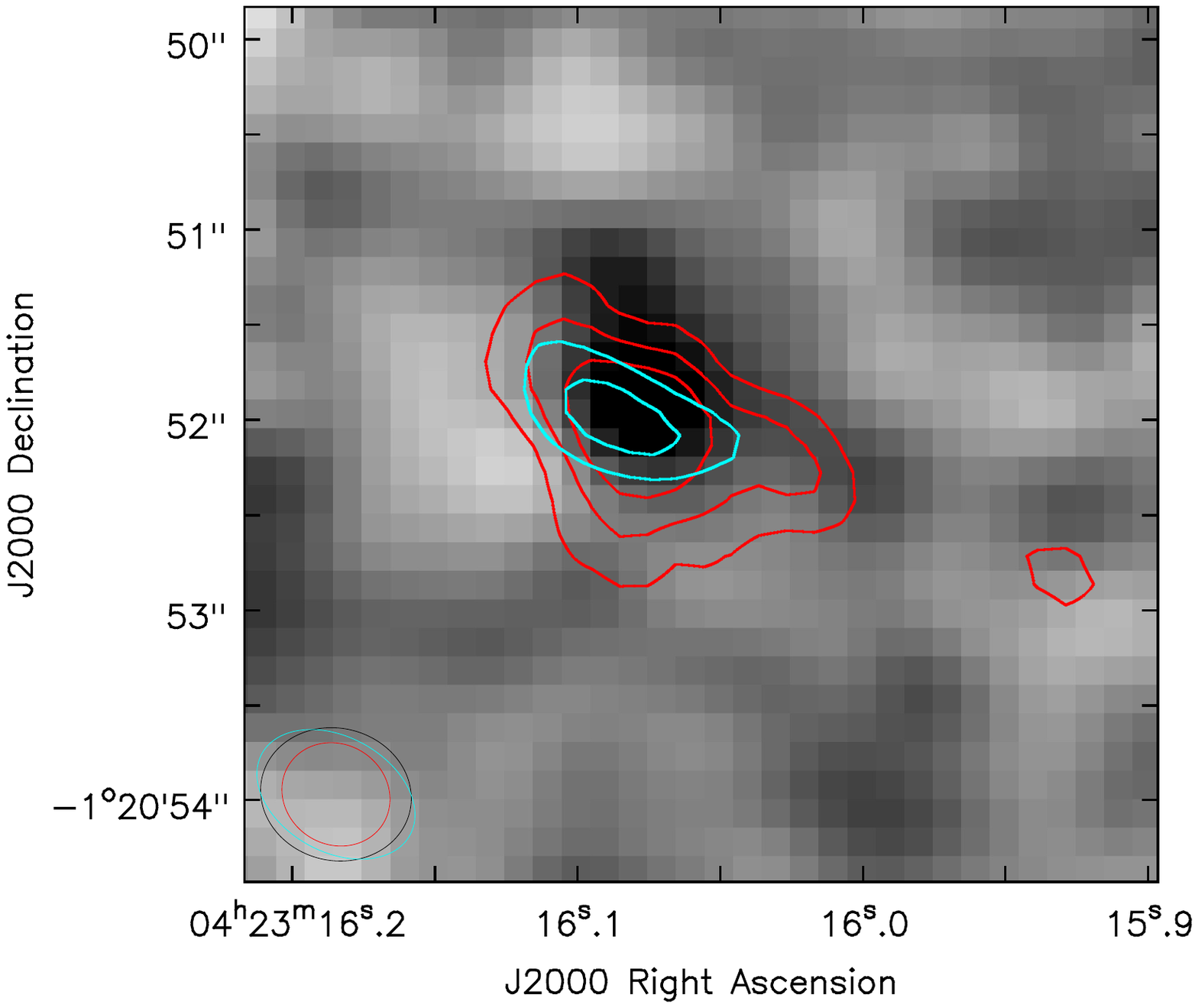}};
    \begin{scope}[x={(image.south east)},y={(image.north west)}]
        \node[align=left] at (0.5, 0.8) {\Large \textbf{1.2mm continuum}};
    \end{scope}
\end{tikzpicture}
\begin{tikzpicture}
    \node[anchor=south west,inner sep=0] (image) at (0,0) {\includegraphics[width = 0.48\linewidth, trim = 10 400 10 10 crop]{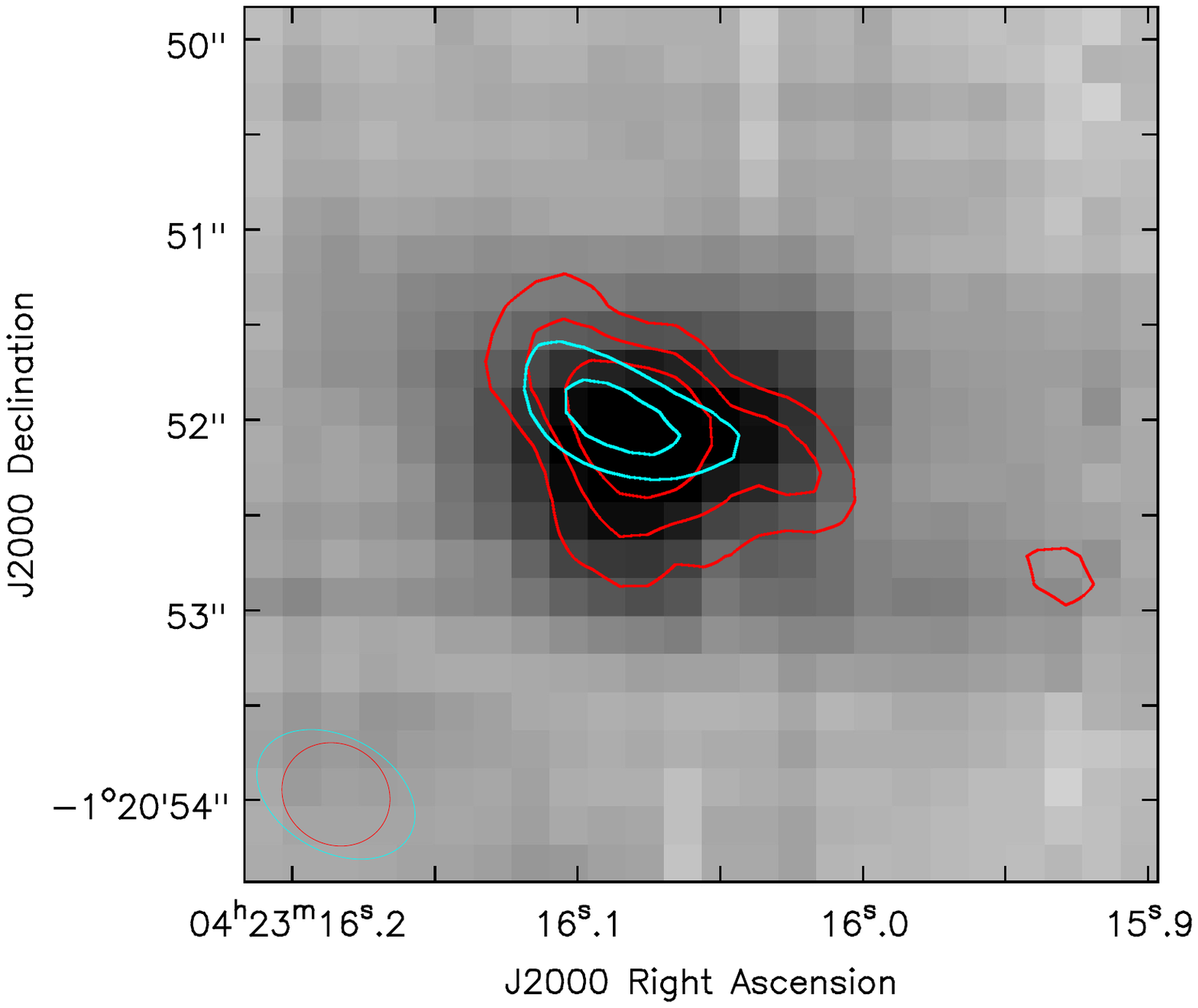}};
    \begin{scope}[x={(image.south east)},y={(image.north west)}]
        \node[align=left] at (0.45, 0.8) {\Large \textbf{optical image}};
    \end{scope}
\end{tikzpicture}
\caption{Zoom in on \GalB\ in Fig.~\ref{Fig.Overview1}. The $1.2$~mm continuum emission is shown in grey scales on the left and the MUSE reconstructed white light image is shown in grey scales on the right. In both panels we show the ALMA $^{12}$CO(2--1) in red contours, ALMA $^{12}$CO (3--2) in cyan contours at $3$, $5$, $7$ sigma of the respective mapdf overlaid. The ellipdfes in the lower left corner show the beam size in the Band~4 and Band~6 observations where the red ellipdfe corresponds to the beam in the $^{12}$CO(2--1) line map, the cyan ellipdfe corresponds to the beam in the $^{12}$CO(3--2) emission line map and the black ellipdfe corresponds to the beam in the $1.2$~mm continuum map. Negative contours are marked with dashed lines, but are not present in this close-up.}% Make the inset again to zoom in on \GalA}
\label{FigMUSEALMA}
\end{figure*}

We aim to detect $^{12}$CO emission from galaxies connected to the Lyman Limit System (LLS) at $z = 0.633$. Thus, we search for emission lines in the ALMA Band~4 and Band~6 image cubes using the Source Finding Application (\texttt{SoFiA})\footnote{\url{httpdf://github.com/SoFiA-Admin/SoFiA}} \mbox{\citep{Serra2015SOFIA}} that incorporates a number of different source detection algorithms to find emission lines in radio data cubes. 
We use the ``Smooth + Clip Finder'', which uses an algorithm developed by \mbox{\citet{Serra2012atlas3D}}. It smooths the data in both spatial and spectral direction using a number of different 3D smoothing kernels. A search of emission lines is performed on each smoothed cube by detecting spaxels above a user-defined threshold. The complete discussion of the relative strengths and weaknesses of all possible source finding strategies is presented by \mbox{\citet{Popping2012comparison}}. We chose the ``Smooth + Clip Finder'' since it offers the highest completeness and reliability for finding sources on a variety of scales. This is appropriate in our case since we do not know a priori whether our sources are spatially resolved. We use twelve smoothing kernels between $0.7 \arcsec$ and $4.2 \arcsec$ in the spatial dimension and between $66$~km~s$^{-1}$  and $990$~km~s$^{-1}$ for the Band~4 data cube and $44$~km~s$^{-1}$  and $660$~km~s$^{-1}$ for the Band~6 data cube in the spectral dimension. The detection threshold is $4\sigma$ for both data cubes.
We require the detection to be within a range of $\pm 2000 \rm{~km~s^{-1}}$ relative to the absorber redshift, because we are targeting $^{12}$CO emission at the absorber redshift. We run the source finder on the non-primary beam corrected cubes to ensure a constant noise level throughout the cube.

%\subsubsection{CO emission lines and Molecular Gas Mass}
\label{SecCOlines}

Using the detection method described above, we find \mbox{$^{12}$CO(2--1)} and $^{12}$CO(3--2) emission at the same position in the Band~4 and Band~6 cubes. We find no other emission lines which are detected in only one of the two data cubes. The emission lines are shown as contours in Fig.~\ref{Fig.Overview1}. The source is named \GalB\ throughout the paper. To our knowledge, this is the first time that multiple $^{12}$CO transitions have been observed from one galaxy associated with a Ly~$\alpha$ absorber. This allows us to investigate the properties of the interstellar-medium in this galaxy.

We determine the size of the emission by fitting a two-dimensional Gaussian function to the integrated intensity mapdf. Here, we report the FWHM along the major axis deconvolved with the beam. The size of the $^{12}$CO(2--1) emission is $1.3 \pm 0.2\arcsec$, the $^{12}$CO(3--2) emission and the $1.2$~mm continuum emission are not resolved.

Before extracting the spectra, we perform a primary beam correction on the image cube using {\sc impbcor}  to account for the primary beam response function. We note that the primary beam correction at the position of our detected $^{12}$CO lines in Band~6 is large because the $^{12}$CO detection lies at $19.4\arcsec$ from the centre of the field, and the Band~6 primary beam width is only $13.5\arcsec$. We determine the redshift of the $^{12}$CO emission lines from the mean of the two frequencies at which the flux reached 50\% of the maximum flux density. 
The emission line spectra are shown in \mbox{Fig.~\ref{FigCOspectra}}.

The flux is measured by integrating the spectra over two times the FWHM indicated by the yellow shaded region shown in Fig.\ \ref{FigCOspectra}. We have subtracted the continuum from the image cube using the task {\sc imcontsub} with a linear fit to the spectrum. We determine the width of the line profiles from the 50\% level of the maximum flux density. The observed line flux densities are converted to line luminosities using the following equation \mbox{\citep{Solomon1992warm}}:
\begin{equation}
L'_{\rm line} = 3.25 \times 10^7 \times S_{\rm line} \Delta v \frac{D_L^2}{(1+z)^3 f_{\rm obs}^2} \rm{K~km~s^{-1}~pc^{2}},
\end{equation}

where $S_{\rm line} \Delta v $ is the observed velocity integrated line flux density in \mbox{Jy km s$^{-1}$}, $D_L$ is the luminosity distance to the galaxy in Mpc, and $f_{\rm obs}$ is the observed line frequency in GHz. A summary of the emission line properties is given in Table \ref{TabCOemissionSummary}. We estimate the error of the integrated line flux to be $5$\% based on the expected accuracy of the flux calibration. The FWHM of the $^{12}$CO emission lines is within the observed range for star-forming dusty galaxies at $z < 0.35$ \citep{Villanueva2017vales}. The projected impact parameter between the $^{12}$CO detection and the quasar sight line is $133$~kpc and the position of the $^{12}$CO detection is aligned with that of one of the galaxies detected in the MUSE cube. We discuss this in more detail in Section \ref{SecSourceMUSE}.

Furthermore, we combine all Band~6 observations from ALMACAL to search for continuum emission from the CO-detected galaxy. Hence, we can compare the total far-infrared luminosity based SFR with the SFR based on the dust-uncorrected [OII] emission line flux. We show in Fig.~\ref{FigMUSEALMA} (left) the Band~6 continuum map with the $^{12}$CO~(2--1) and $^{12}$CO~(3--2) emission line contours overplotted. We exclude any contamination from the $^{12}$CO(3-2) emission line by excluding the datasets covering the relevant frequency. The total flux density at $1.2$~mm is $(0.8 \pm0.2)$~mJy and the peak intensity is $(0.56 \pm 0.09)$~mJy~beam$^{-1}$.% We find that the total $1.2$~mm flux is consistent with the measured SFR and molecular gas mass in this galaxy regarding Eq.(15) in \citet{Scoville2016ISMmasses}, which yields an expected $1.2$~mm flux of $0.1 - 0.9$~mJy.}

Finally, we explore the possibility to observe molecular absorption lines towards the quasar in our Band~4 and Band~6 observations. The $3\sigma$ detection limit for the integrated optical depth is $0.01$~km~s$^{-1}$ and $0.003$~km~s$^{-1}$ for $^{12}$CO~(2--1) and $^{12}$CO~(3--2), respectively. We note, however, that the velocity resolution is very coarse and might be not sufficient to detect absorption lines.

\subsection{MUSE Source detection and Flux Measurement}
\label{SecSourceMUSE}

\begin{table*}
\caption{Observed properties of the galaxies from the MUSE observation and the broad-band imaging.}
\label{TabPhotometry}
\label{TabEmissionLines}
\begin{tabular}{l l c c c c}
\hline
Name & & \GalA~& \GalB\ & \GalC\ & \GalD\\
\hline
Ra (J2000) & & 04 23 16.18 & 04 23 16.07 & 04 23 15.54 & 04 23 15.30\\
Dec (J2000) & & $-$01 20 46.9 &  $-$01 20 52.1 & $-$01 20 54.0 &  $-$01 21 03.7\\
$\theta$ [\arcsec] & & $14.9$ & $19.4$ &  $21.3$ & $31.6$\\
$\theta$ [kpc] & & $102$ & $133$ & $146$ & $216$\\
$z$  & & $0.6332 \pm 0.0005$ & $0.6331 \pm 0.0005$ & $0.6338 \pm 0.0003$ & $0.6323 \pm 0.0005$ \\
Line Width & $\sigma$([OII]) & $170 \pm 10$ & $220 \pm 10$ & $170 \pm 10$ & $140 \pm 10$ \\
& FWHM([OII]) & $400 \pm 20$ & $520 \pm 30$ & $390 \pm 30$ & $330 \pm 30$\\
 & $\sigma$([H~$\beta$]) & $120 \pm 50$ & - & $120 \pm 40$ & - \\
 & FWHM([H~$\beta$]) & $290 \pm 110$ & - & $290 \pm 90$ & - \\
  & $\sigma$([OIII]5007\AA) & $150 \pm 20$ & $170 \pm 30$ & $160 \pm 40$ & $70 \pm 20$\\
  & FWHM([OIII]5007\AA) & $360 \pm 50$ & $400 \pm 70$ & $380 \pm 90$ & $160 \pm 40$\\
Line Fluxes & \\
$[10^{-17}$ erg s$^{-1}$ cm$^{-2}]$& [OII] & $5.3 \pm 0.5$ & $11.0 \pm 1.1$ & $3.6 \pm 0.4$ & $3.0 \pm 0.3$\\
& H~$\beta$ & $1.7\pm 0.2$ & $<0.9$ & $1.7 \pm 0.2$	& $<2$\\
& [OIII] 4959 & $<0.9$ & $<1$ & $<0.6$ & $<1$\\
& [OIII] 5007 & $2.1 \pm 0.2$ & $2.9 \pm 0.3$ & $1.3 \pm 0.1$ & $2.2 \pm 0.2$\\
$B$ [mag] & & $23.8 \pm 0.4$ & $25.2 \pm 0.5$ & $24.6 \pm 0.3$ & $<25.82$\\
$R$ [mag] & & $22.10 \pm 0.18$ & $21.68 \pm 0.18$ & $22.72 \pm 0.15$ & $23.75 \pm 0.17$\\
$I$ [mag] & & $21.0 \pm 0.3$ & $20.0 \pm 0.2$ & $21.8 \pm 0.2$ & $<25.15$ \\
$J$ [mag] & & $20.43 \pm 0.12$ & $19.64 \pm 0.09$ & $20.90 \pm 0.13$ & $21.67 \pm 0.11$\\
$H$ [mag] & & $19.83 \pm 0.19$ & $18.91 \pm 0.12$ & $20.3 \pm 0.2$ & $20.90 \pm 0.19$\\
$K$ [mag] & & $19.87 \pm 0.13$ & $18.78 \pm 0.08$ & $20.34 \pm 0.15$ & $20.80 \pm 0.11$\\
\hline
\end{tabular}
\begin{minipage}{\linewidth}
Note: $\theta$ is the impact parameter given here in \arcsec\ and kpc.
\end{minipage}
\end{table*}

%\subsubsection{Source detection}

\begin{figure*}
\includegraphics[width = \linewidth]{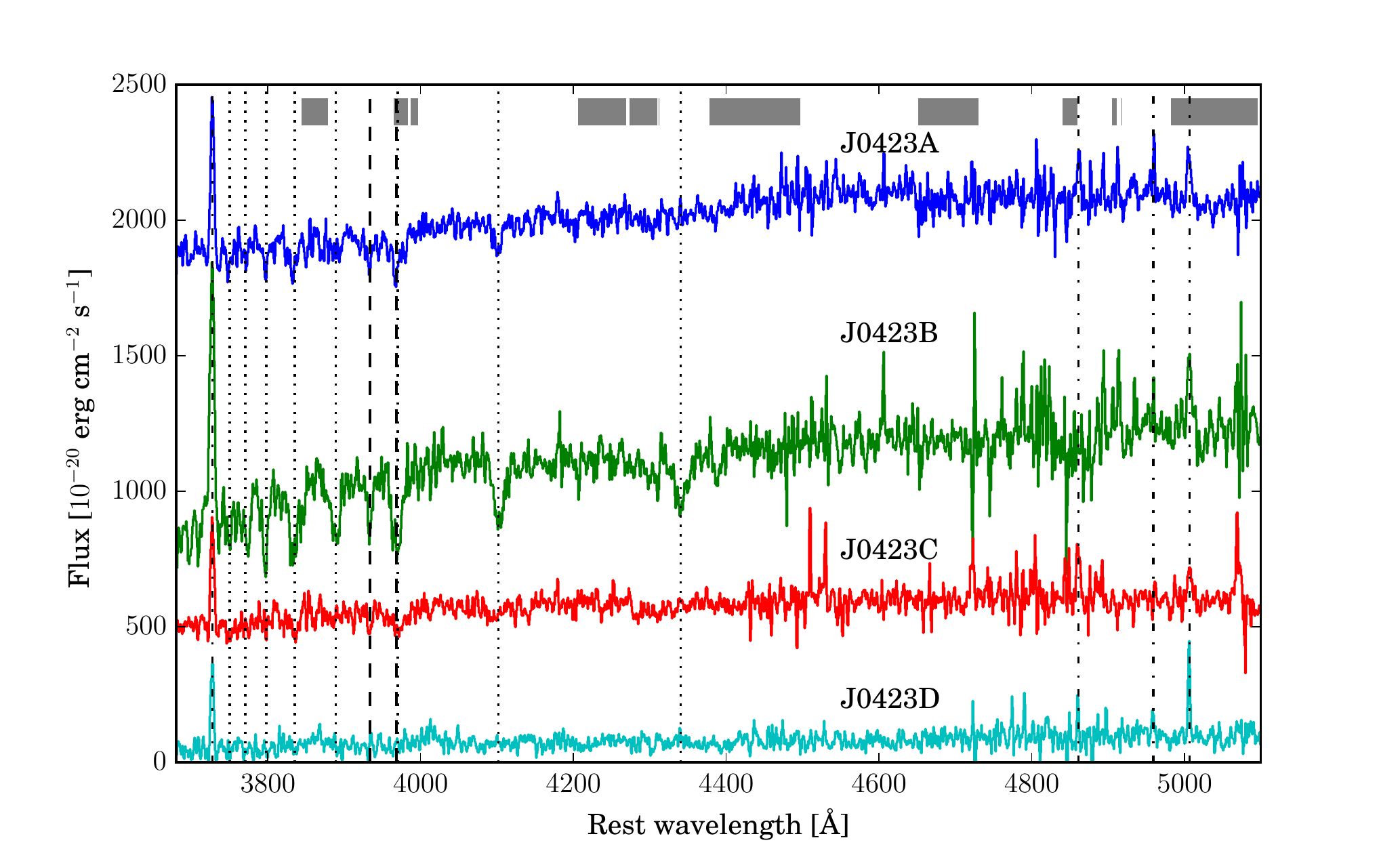}
\caption{Rest frame optical spectra of the four group members identified from our MUSE observation (offset by an arbitrary number for clarity). The spectra are shifted to rest wavelength using the redshift determined from the emission lines quoted in Table \ref{TabEmissionLines}. The grey area marks the regions possibly contaminated by telluric absorption and emission. The strongest emission line in all spectra is [OII] at $\lambda = 3727$~\AA. The dashed-dotted line marks with increasing wavelength the [OII], H~$\beta$, [OIII]$4959$\AA, and [OIII]$5007$\AA\ emission line. The dotted lines mark the Balmer absorption lines from H~$\gamma$ to H~12 with decreasing wavelength. The dashed lines mark the Ca~H\&K absorption lines. We note, that the features at the [OIII]$4959$\AA\ and H~$\beta$ wavelength in \GalD\ as well as the narrow feature at the [OIII]$4959$\AA\ wavelength in \GalA\ are marginal given the SNR of the spectrum.}
\label{FigFullSpectra}
\end{figure*}

We blindly search for line and continuum emission from galaxies in our MUSE cube. To this end, we search for emission line and continuum sources using the MUSELET source finding algorithm included in MPDAF\footnote{http://mpdaf.readthedocs.io/en/latest/}. MUSELET runs SExtractor \mbox{\citep{Bertin1996SExtractor}} on each $1.25$~\AA-wide slice of the cube to find any source of emission. We detect $50$ emission line sources and continuum sources in the MUSE cube with this method. The galaxies at the absorber redshift ($z = 0.633$) are selected based on two criteria: 1) searching for emission line sources, and 2) by searching for the $4000$~\AA\ break at the rest frame of the absorber in all $50$ spectra. We find four emission line sources at the absorber redshift, but we do not detect continuum sources at the absorber redshift. The full spectra of the four emission line galaxies are shown in Fig.~\ref{FigFullSpectra}. In Fig.~\ref{Fig.Overview1} we mark the positions of these galaxies: \GalA, \GalB, \GalC\ and \GalD.

%\begin{figure}
%\begin{tikzpicture}
%    \node[anchor=south west,inner sep=0] (image) at (0,0) {\includegraphics[width = \linewidth, trim=50 180 50 180,clip]{[OII]NarrowBandMinusContPositions.pdf}};
%    \begin{scope}[x={(image.south east)},y={(image.north west)}]
%        \draw [ultra thick] (0.1,0.1) -- node[below]{\Large 50 kpc} (0.21,0.1);
%    \end{scope}
%\end{tikzpicture}
%\caption{pdfeudo-narrow-band MUSE image of the [OII] line at the absorber redshift. We integrate over a velocity range of 865 km s$^{-1}$ covering the [OII] emission from all four detected galaxies from our MUSE observation. The continuum averaged over $\pm 1000 - 10000 \: \text{km}\: \text{s}^{-1}$ is subtracted.  
%The galaxies at the absorber redshift are marked with green circles and are named \GalA, \GalB, \GalC, and \GalD\ hereafter. The other sources seen in this figure are three stars and the quasar in the centre. The scale indicates 50~kpc at the absorber redshift.}
%\label{FigMUSELETdetections}
%\label{FigOIILineMap}
%\end{figure}

%\subsubsection{Emission lines}

We determine the redshift of the galaxies by fitting a Gaussian to the detected emission lines. Since we detect multiple lines for all galaxies, we determine the redshift of each galaxy as the median redshift from the Gaussian fit to the detected lines. The respective redshifts and emission line fluxes are shown in \mbox{Table \ref{TabEmissionLines}}. We do not detect [OIII]$4959$\AA\ for any of our galaxies and therefore give only upper limits on the line flux. This is consistent with the expected line-flux assuming a doublet flux-ratio of $3$ \mbox{\citep{Storey2000theoretical}}. Furthermore, we detect H~$\beta$ in only two of the galaxies. The given upper limits are calculated from a maximum emission of $3\sigma$ and the FWHM measured from the sky emission lines corresponding to $3$~\AA. The line width of the detected lines determined from the Gaussian fitting is listed in Table \ref{TabEmissionLines}.

In Fig.\ \ref{FigMUSEALMA}~(right) we show a comparison of the ALMA collapdfed $^{12}$CO emission line map and the MUSE pdfeudo-white-light image. It can be seen that the $^{12}$CO gas is coincident with the optical position of galaxy \GalB. No emission from the other MUSE-detected galaxies is found in the ALMA cubes. Furthermore, comparing the FWHM of the [OIII] lines and the $^{12}$CO emission lines, we find that the molecular gas disk seems slightly more extended than the ionized gas.

\subsection{Broad-band Photometry and Stellar Mass}
\label{SecBroadBandPhot}

We estimate the stellar mass of the four galaxies, to infer the molecular gas mass ratio of \GalB\ and determine the total dynamical mass of the group. For this purpose, we use the broad-band imaging from \mbox{\citet{Rao2011groundbased}} to measure the broad-band photometry which we use as input for the spectral energy distribution (SED) fitting. The apparent magnitudes of the detected galaxies are determined using SExtractor and are listed in Table \ref{TabPhotometry}. 

This broad-band photometry is used as input for the SED fitting with Le Phare\footnote{ \url{http://www.cfht.hawaii.edu/~arnouts/lephare.html}} \mbox{\citep{Arnouts1999measuring, Ilbert2006accurate}} to estimate the stellar masses of the galaxies. To perform the SED fitting the algorithm compares the observed colours with the ones predicted from a set of template SEDs. We fix the redshift for the SED fitting to the spectroscopic redshift determined from our MUSE spectra described in Section \ref{SecSourceMUSE}. The template SEDs are convolved with standard $B$, $R$, $I$, $J$, $H$, and $K$ filter functions and a $\chi^2$ minimization is performed. %The SED fitting yields the total IR luminosity, the stellar mass, SFR, and the age of the stellar population. However, our broad-band photometry does not cover a wide wavelength range and therefore only the stellar mass is well constrained by the fitting.}
We quote the stellar mass of the galaxies in Table~\ref{TabPhysParam}. %The other parameters are not as reliable and show large errorbars in the fitting. 
Comparing our stellar masses to the galaxy stellar mass function at redshift $0.4<z<0.6$ and $0.6<z<0.8$   for which \mbox{\citet{Drory2009bimodal}} report a log(M$^{\star}$) of $10^{10.91 - 10.95}$, we find that the CO-detected, most massive galaxy \GalB\ has a stellar mass of almost $2$~M$^{\star}$, while the other galaxies \GalA, \GalC\ and \GalD\ have stellar masses below M$^{\star}$.\\

\subsection{Molecular Gas Mass}
\label{SubsecMolGas}

Until now the molecular gas content of galaxies associated with intervening Ly~$\alpha$ absorbers has been determined only once \citep{Neeleman2016first}. We combine the molecular gas mass with the stellar mass from the SED fitting to determine a census of the molecular gas and stars in galaxies associated with intervening absorbers. In the following, we determine the molecular gas mass using the two $^{12}$CO emission lines. We use the following conversion:

\begin{equation}
M_{\rm mol} = \alpha_{\rm CO} \times L'_{\rm CO(1-0)},
\end{equation}

where $L'_{\rm CO(1-0)}$ is the luminosity of the $^{12}$CO(1-0) emission line and the conversion factor $\alpha_{\rm CO}$ depends on the temperature, density, metallicity, and column density of the interstellar medium.

To apply the above mentioned conversion to the total molecular gas mass, we need to convert the $^{12}$CO(2--1) line flux to the $^{12}$CO(1-0) line flux using a suitable conversion factor. The ratio of $L'_{\rm CO(3-2)}/L'_{\rm CO(2-1)}$ is $0.63 \pm 0.09$ suggesting a $^{12}$CO excitation ladder similar to that of a Luminous Infrared Galaxy (LIRG) \mbox{\citep{Papadopoulos2012molecular1}}. We note that this value is also consistent with the expected ratio of $0.54$ for Milky Way-type galaxies \citep{Carilli2013cool}. However, based on further evidence presented in Sec.~\ref{SubsecJ0423B-LIRG} we believe that the LIRG-type conversion factor is more appropriate. Therefore, we adopt a conversion factor of \mbox{$L'_{\rm CO(2-1)}/L'_{\rm CO(1-0)} = 0.9$}, yielding an $L'_{\rm CO(1-0)}$ of \mbox{$(2.1 \pm 0.1) \times 10^{10}$ K km s$^{-1}$ pc$^{2}$}. Furthermore, we use $\alpha_{\rm CO} = 0.6 \rm{M}_{\sun}\: (\rm{K\: km\: s^{-1} pc^2})^{-1}$, appropriate for LIRGs derived by \mbox{\citet{Papadopoulos2012molecular2}} including a factor of 1.36 to account for the presence of helium. This yields a molecular gas mass of \mbox{$M_{\rm mol} = (1.27 \pm 0.07) \times 10^{10} \: \rm{M}_{\sun}$.} To indicate the uncertainty in the derived H$_{2}$ mass, we also apply a conversion factor of $\alpha_{\rm CO} = 4.3 \rm{M}_{\sun}\: (\rm{K \: km\: s^{-1} pc^2})^{-1}$ typical for Milky Way-type galaxies and including a factor of 1.36 to account for the presence of helium \mbox{\citep{Bolatto2013co}}.
This yields a molecular gas mass $9 \times 10^{10}\: \rm{M}_{\sun}$ and so we adopt a molecular gas mass of $1 - 9 \times 10^{10}\: \rm{M}_{\sun}$.\\

We have also extracted spectra from our ALMA cubes at the positions of \GalA\ and \GalC, but we could not find any clear sign of emission. For these two galaxies, we calculate upper limits for $L'_{\rm CO(2-1)}$ based on the $3\sigma$ noise level and assuming a width of the line of $500 \rm{km\; s ^{-1}}$ based on the $w_{50}$ measured from the $^{12}$CO detections for \GalB. This yields an upper limit for the integrated flux of \mbox{$S_{\rm int} = 0.4 \rm{ \;Jy\;km\;s^{-1}}$}, assuming that the flux is evenly distributed over the full width of the profile. Therefore, the upper limit on the line luminosity of $L'_{\rm CO(2-1)}$ is $2 \times 10^{9}  \rm{\;K\;km\;s^{-1}\;pc^{2}}$. We assume a Milky Way-type $L'_{\rm CO(2-1)}/L'_{\rm CO(1-0)}$ emission line ratio of $0.5$ \citep{Carilli2013cool} and a Milky Way-type \mbox{CO-to-H$_2$} conversion factor of \mbox{$\alpha_{\rm CO} = 4.3  \rm{M}_{\sun}\:\rm{(K\;km\;s^{-1}\;pc^2)^{-1}}$} including a factor of 1.36 to account for the presence of helium  \mbox{\citep{Bolatto2013co}}. This yields an upper limit for the molecular gas mass of $M_{\rm mol}< 1.7\times 10^{10}$~M$_{\sun}$ for \GalA\ and \GalC.

\subsection{Star-Formation Rates}
\label{SubsecSFR}

We determine the SFR of the CO-detected galaxy \GalB\ to test whether it is comparable with the identification of a LIRG based on the CO line flux ratios.

We estimate the SFR of \GalB\ based on the $1.2$~mm flux. We use a set of template spectra for starburst galaxies from \citet{Polletta2007spectral} which we scale to the observed $1.2$~mm flux converted to the rest wavelength. The fitted spectrum is then integrated in the wavelength range from $8$~-~$1000 \:\mu \text{m}$ to obtain the total far-infrared luminosity $L_{\rm FIR}$. This is converted to the SFR using the \citet{Kennicutt1998starformation} relation:

\begin{equation}
SFR_{\rm FIR} = 4.5 \times 10^{-44} L_{\rm FIR}
\end{equation}

based on this, we estimate a SFR of \mbox{$(50 \pm 10)$~M$_{\sun}$~yr$^{-1}$}. We note, however, that the uncertainty in this calculation is high since we use the median SFR from a set of template spectra which we scale to the flux at $1.2$~mm.

Furthermore, we derive the SFRs for the remaining three emission line galaxies based on the [OII] emission line using the relation given by \mbox{\citet{Kennicutt1998starformation}}, which includes a dust correction:

\begin{equation}
SFR_{\rm [OII]} = (1.4 \pm 0.4) \times 10^{-41} L([\rm{OII}]) \times 10^{0.4 A_{\rm [OII]}} .
\end{equation}

where $A_{\rm [OII]}$ is the extinction of the [OII] emission line. We determine the dust correction based on the stellar mass dependent $A_{\rm H\; \alpha}$ given by \citet{Garn2010predicting} at $z \sim 0.1$, which is found to be valid up to $z \sim 1.5$ \citep{Sobral2012star}. To convert the $A_{\rm H\; \alpha}$ to $A_{\rm [OII]}$ we use the prescription given by \citet{Calzetti1997reddening}. The resulting dust-corrected SFRs are given in Table \ref{TabPhysParam}. 

To test the identification of \GalB\ being a LIRG, we calculate the dust-uncorrected  SFR based on the [OII] emission line for a comparison with the infrared based SFR. We derive a SFR of \mbox{$2.7 \pm 0.3$~M$_{\sun}$~yr$^{-1}$} from the [OII] emission. This is $13$~-~$25$ times lower than the SFR inferred from the far-infrared luminosity. This is in the lower envelope of the population of very luminous infrared galaxies studied by \citep{Poggianti2000optical}. This is a further piece of evidence that \GalB\ is a LIRG.

%We find that the $^{12}$CO-detected galaxy has the highest SFR of $37$~M$_{\sun}$~yr$^{-1}$. This high SFR is consistent with the identification as a LIRG based on the $^{12}$CO-line flux ratio.

We determine the limiting SFR for our MUSE observations based on the detection limit of the [OII] line to infer the maximum SFR of potential galaxy candidates below our detection limit. The mean r.m.s. noise in the MUSE cube around the [OII] emission line is measured from the cube with a velocity width of $865$~km~s$^{-1}$. We perform a $3\sigma$ clipping on this cube and another $3\sigma$ clipping on the clipped cube to remove any contribution from actual emission. The mean r.m.s. noise is then determined from this emission-free cube. Furthermore, we assume, that the minimum size of a galaxy is given by the seeing of $0.7\arcsec$, which corresponds to $4.8$~kpc at the absorber redshift and the FWHM of the line is at least $3$~\AA. The resulting [OII] flux would be $0.3 \times 10^{-17}$ erg~s$^{-1}$~cm $^{-2}$. This converts into a 3$\sigma$ limiting SFR of $0.2$~M$_{\sun}$~yr$^{-1}$. Additionally, we determine the limiting SFR at the quasar position. Therefore, we extract a $2\arcsec \times 2\arcsec$ wide cube at the position of the quasar with a width of $865$~km~s$^{-1}$ centred on the expected [OII] emission line covering all the observed [OII] emission. The emission from the quasar is determined by a linear fit and removed from the data cube. The limiting SFR at the position of the quasar is calculated from the noise in the continuum subtracted cube and assuming the same emission properties as described above. This yields a non-dust-corrected limiting SFR of $0.3\text{ M}_{\sun}\text{ yr}^{-1}$.

\begin{table*}
\begin{minipage}{\linewidth}
\caption{Summary of the physical properties of the galaxies at the absorber redshift based on our MUSE data and the SED fitting.}
\label{TabPhysParam}
\centering
\begin{tabular}{lllllllll}
\hline
Name & $z$ & SFR  & $\log M_{\star}$ & $12$ + log(O/H)\\
& & [M$_{\sun}$~y$r^{-1}$] &  [M$_{\sun}$] & \\
\hline
\GalA\ 	& $0.63317 \pm 0.00048$ & $7.5 \pm 0.6$ & $10.4 \pm 0.2$ 	& $8.80 \pm 0.10^a$ \\%& 7.95 \pm 0.14 & \\%[$9.01 \pm 0.04$]\\
\GalB\ 	& $0.63312 \pm  0.00048$ & $50 \pm 10$ &  $11.2 \pm 0.1$ &  $9.1 \pm 0.9^b$\\
\GalC\ 		& $0.63376 \pm 0.00032$ & $4.6 \pm 0.5$  & $10.3 \pm 0.2$ & $8.94 \pm 0.06^a$  \\%& 7.71 \pm 0.12\\
\GalD\ 		&  $0.63229 \pm 0.00048$ & $3.2 \pm 0.5$  & $10.2 \pm 0.2$  & $9.0 \pm 0.9^b$\\
\hline
\end{tabular}
\end{minipage}
\begin{minipage}{\linewidth}
Note: The SFRs of \GalA, \GalC, \GalD\ reported in this table are based on the [OII] emission line and are dust corrected. The SFR of \GalB\ is derived from $L_{\rm FIR}$. The metallicity $^a$ is determined from the emission lines and the metallicity $^b$ is derived from the mass-metallicity relation.
\end{minipage}
\end{table*}

\section{Discussion}

\subsection{The Nature of the galaxy J0423B}
\label{SubsecJ0423B-LIRG}

Galaxy \GalB\ is detected in $^{12}$CO(2--1) and the $^{12}$CO(3--2) as well as our MUSE observations. Here we collect all this information and discuss the nature of this galaxy. In \S~\ref{SubsecMolGas} it is shown that the ratio of the $^{12}$CO(2--1) and the $^{12}$CO(3--2) emission line is consistent with \GalB\ being a LIRG. Furthermore, the SFR based on $L_{\rm FIR}$ reported in \S~\ref{SubsecSFR} is $50 \pm 10~\text{M} _{\sun}\; \text{yr}^{-1} $. We compare this to the SFR from the dust-uncorrected [OII] line flux, which is $13$~-~$25$ times lower. This is in the lower envelope of the very luminous infrared galaxy population studied by \citet{Poggianti2000optical}. Combining these pieces of evidence we conclude that \GalB\ is a LIRG.

\begin{figure}
\includegraphics[width = \linewidth]{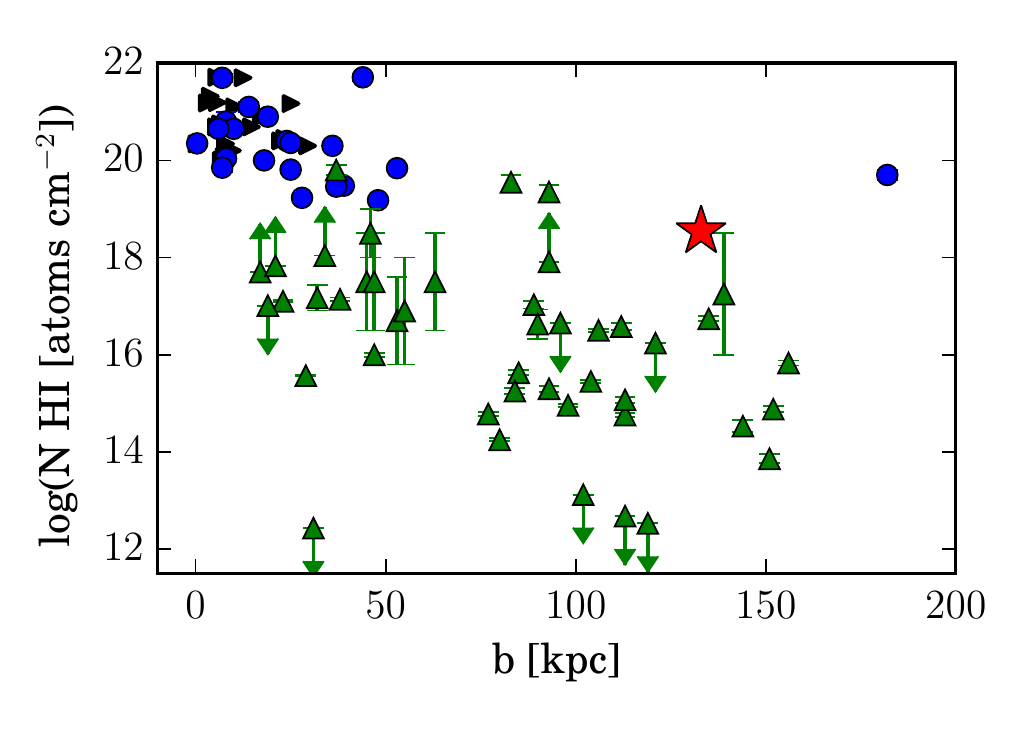}
\caption{Column density of the intervening absorber as a function of impact parameter from \GalB\ (red star), DLAs from \mbox{\citet{Zwaan2005reconciling}} (black triangles) and \mbox{\citet{Peroux2011sinfoni}} (blue circles) and subDLAs and LLS from the COS halos survey \mbox{\citep{Prochaska2017cos}} (green triangles).}
\label{FigColumnImpactComp}
\end{figure}

\begin{figure}
\includegraphics[width = 0.8 \linewidth]{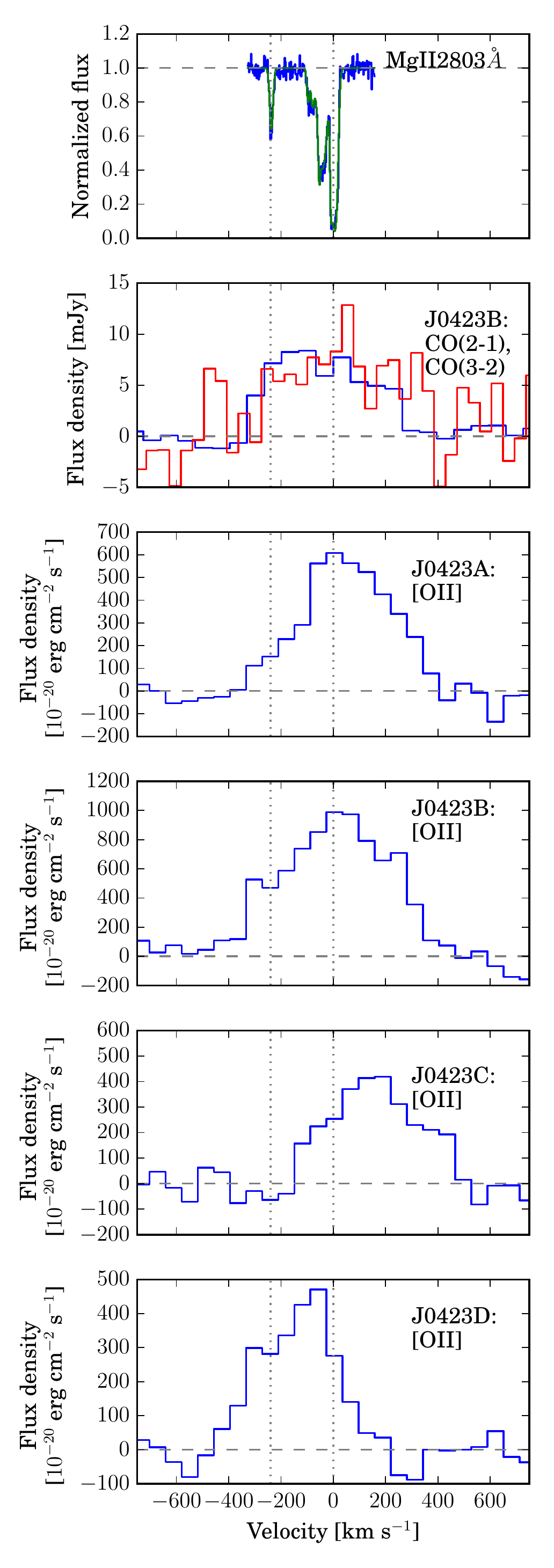}
\caption{Comparison of all detected systems from top to bottom: MgII absorption in the quasar spectrum from the HIRES spectrum, $^{12}$CO(2--1) (blue) and $^{12}$CO(3--2) (red) from \GalB\ from ALMACAL, the continuum subtracted [OII]~emission line spectra from \GalA, \GalB, \GalC\ and \GalD\ all plotted relative to the absorber redshift. We stress that the width of the $^{12}$CO and [OII] emission lines cannot be directly compared in this plot since the [OII] emission line doublet is not resolved in the MUSE observations.}
\label{FigAllComp}
\end{figure}

\subsection{Assessing the Possibility of a Misidentification}

The fact that we only detect galaxies at large impact parameters from the quasar sight-line raises the question whether an undetected galaxy hiding in the bright quasar light could be responsible for the absorption. Interestingly, \mbox{\citet{Churchill1996spatial}} report a galaxy at an impact parameter of $2.1\arcsec$ or $14.6$~kpc and \mbox{\citet{Rao2011groundbased}} report another galaxy at an impact parameter of $3.6\arcsec$ or $25$~kpc. We do not detect either of these galaxies in our data. In addition, these objects are not seen in quasar pdfF-subtracted archival HST/NICMOS images (PID 7451).
We determine the non-dust-corrected limiting SFR of $0.2\text{ M}_{\sun} \text{ yr}^{-1}$ over the full field of view and $0.3 \text{ M}_{\sun} \text{ yr}^{-1}$ in a $2\arcsec \times 2 \arcsec $ wide field centred on the position of the quasar. We take this as the upper limit on the SFR of an undetected galaxy within the quasar point spread function.\\

We compare the HI column density of the absorber and the impact parameter of the most massive group member with the column density~-~impact parameters observed for DLAs \mbox{\citep{Zwaan2005reconciling, Peroux2011sinfoni}} and subDLAs and LLS from the COS Halos survey \mbox{\citep{Prochaska2017cos}} in Fig.~\ref{FigColumnImpactComp}. Assuming that \GalB\ can be uniquely associated with the Ly~$\alpha$ absorption line, we see it is in the upper envelope of the relation between impact parameter and column density, but not atypical compared to other identified galaxy-absorption line pairs from the literature. This lessens the immediate need for another galaxy at a smaller impact parameter.

\subsection{The Nature of the Absorbing Gas}
%\subsection{Could the Ly~$\alpha$ Absorption be Tracing an Outflow?}

Here, we discuss the possibility that the absorbing gas is tracing an outflow or intra group medium.

We consider whether the gas seen in absorption could be related to an outflow from one of the identified galaxies. In this galaxy group an outflow is most likely starburst-driven and would, therefore, originate from \GalB. This is further supported by the morpho-kinematic properties of the galaxies studied in \S~\ref{SecGalfit}. If we assume that the outflow of gas from galaxies is most efficient along the minor axis theoretically predicted by \mbox{\citet{Keres2009galaxies, Stewart2011orbiting}} and empirically motivated by recent studies {\citep[e.g.][]{Bordoloi2011radial, Bouche2012enriched, Schroetter2015sinfoni} the best candidate having an outflow that could be probed by the absorption is \GalB. Furthermore, we assume a constant typical velocity of a galactic wind in a star-forming galaxy of $300$~km~s$^{-1}$\mbox{\citep{Veilleux2005galactic}}. This yields a travel time of $\sim 430$~Myr from \GalB\ to the position where we see the absorption.
%The gas we probe in absorption could be an outflow since the line of sight is close to the minor axis \GalB. If we assume a constant typical velocity of a galactic wind in a star-forming galaxy of 300~km~s$^{-1}$\mbox{\citep{Veilleux2005galactic}} it would take $\sim$ 430~Myr to travel from \GalB\ to the position where we see the absorption.}
%Regarding the $^{12}$CO emission line ratios, which suggest LIRG-type ISM conditions and a SFR~$\sim 37$~M$_{\sun}$~yr$^{-1}$, it is plausible that we observe a starburst-driven outflow. 
Recently, first attempts in theoretical modelling were made to understand how a cool outflow can be produced \mbox{\citep{Richings2017origin}}.

%\subsection{Could the Absorption Trace Intra-group Gas?}
Another explanation for the gas seen in absorption could be that it is tracing the intra-group medium.
The MUSE observations revealed four galaxies at the absorber redshift, which is an over-density compared to the total redshift range. We expect based on the [OII] luminosity function given by \citet{Ly2007luminosity} 0.05 galaxies above the limiting SFR within the MUSE field of view in a redshift range covered by the galaxy group. Therefore, we finally consider whether the gas seen in absorption against the background quasar is probing the CGM of the group of galaxies that we identify to be close in redshift space and in projected separation. We compare the alignment of the galaxies and the absorber in velocity space. In Fig.~\ref{FigAllComp} we show the position of the absorber, the $^{12}$CO detections from the  ALMA observations and the ionized gas detected with MUSE in velocity space with respect to the absorber redshift. We find that all galaxies align well with the absorber, where \GalC\ is redshifted by $200$~km~s$^{-1}$ and \GalD\ is blueshifted by $-250$~km~s$^{-1}$ with respect to the absorber systemic velocity. This could indicate that the quasar absorption line probes the intra-group gas rather than the CGM of a single galaxy. Intervening absorbers are often found to trace intra-group gas at low redshift \citep{Whiting2006mos, Kacprzak2010galaxy, Gauthier2013ultra, Rahmani2017observational, Bielby2017probing, Fumagalli2017witnessing, Peroux2017nature}. In these cases the velocity width of the absorption ranges from less than $100$~km~s$^{-1}$ to more than $600$~km~s$^{-1}$. At this stage, we cannot distinguish intra-group gas from CGM based on this property.

The spatial extent of the CGM in group galaxies is reported to be more than $140$~kpc and the intra-group gas can be even more extended \mbox{\citep{Bordoloi2011radial}}.

As an illustration, we compute an estimate of the dynamical mass of the group and the radius of the zero-velocity surface. To calculate the dynamical mass, we use the `projected mass estimator' \mbox{\citep{Heisler1985estimating}} defined by:

\begin{equation}
M_{\rm PM} = \frac{32/\pi}{\text{G}(N_{\rm m} - 1.5)}\sum_i v_i^2 r_i,
\end{equation}

where G is the gravitational constant, $N_{\rm m}$ is the number of observed group members, $v_i$ is the velocity of the group member $i$ with respect to the group mean velocity and $r_i$ is the projected distance of the group member $i$ from the group centre. Here we define the most massive galaxy as the group centre since we cannot be sure that all group members are observed. This yields a dynamical mass of the group of $10^{12.9} \text{M}_{\sun}$. Furthermore, we compute the radius of the zero-velocity surface, $R_0$, defined as the surface beyond which galaxies participate in the Hubble expansion \mbox{\citep{Sandage1986redshift}}. It is derived as follows:

\begin{equation}
R_0 = \left( \frac{8 G T^2}{\pi^2} M_{\rm dyn} \right)^{1/3},
\end{equation}

where $T$ is the age of the group. We assume that the age of the group is $7.5$~Gyr, which is the age of the universe at the redshift of the group. This yields a radius of the zero-velocity surface of $1.2$~Mpc. The gas that is probed by the Ly~$\alpha$ absorption is well within this radius.

Furthermore, it is well known from nearby interacting groupdf that the group galaxies can be embedded in a large-scale atomic gas reservoir \citep{Yun1994highres}.

%\subsection{Other explanations}

%\begin{itemize}
%\item Merger
%\item intra group medium / IGM filament
%\begin{itemize}
%\item metallicity of the absorbing gas is -1.16, the gas must have been reprocessed
%\item see Prochaska et al. 2011 (filament)
%\item Peroux et al. 2016 (intra group gas)
%\item Bielby et al 2016 (intra group gas, group within 200kpc, closest 49kpc, column density $10^{19.9}$ atoms/cm$^2$)
%\end{itemize}
%\end{itemize}

\section{Summary and Conclusion}

%\subsection{Combining Molecular Gas, HII Regions and Neutral Gas}

In this case study of a Lyman Limit System at $z = 0.633$, we show the power of combining observations of the cool ionized gas with MUSE and ALMA observations of the $^{12}$CO emission lines tracing the molecular gas in galaxies connected with a Ly~$\alpha$ absorber. We measure two rotational transitions of the $^{12}$CO line in emission, corresponding to a galaxy at the same redshift as the Ly~$\alpha$ absorption line. This allows us to determine the molecular gas mass, as well as to put constraints on the ISM properties of this galaxy. We can efficiently identify star-forming galaxies at the absorber redshift using additional optical IFU observations. With these observations, we can probe the cool ionized gas in the galaxies and study their morpho-kinematics. In combination with the observations of the neutral gas probed in absorption, the observations of the cool ionized and molecular gas offer a complete census of the cool gas, which is needed to gain a better understanding of what types of galaxies are probed by Ly~$\alpha$ absorbers.\\

In the particular system presented in this study we find a group of massive galaxies with masses of M$_{\star} = 10^{10.3 - 11.2}\, \text{M}_{\sun}$ connected to a LLS with $\log ( N(\rm{HI}) / \: \text{atoms} \: \text{cm}^{-2}) = 18.54^{+0.07}_{-0.10}$ at a redshift of $z=0.633$. The impact parameter between the quasar sightline and the closest galaxy is $102$~kpc and $133$~kpc for the most massive galaxy. No other galaxy is detected closer to the quasar down to a dust-uncorrected limiting SFR of $0.2 \text{ M}_{\sun} \text{ yr}^{-1}$ in the field and $0.3 \text{ M}_{\sun} \text{ yr}^{-1}$ at the quasar position. For the three lower mass galaxies, we find a dust corrected SFR of $3$~M$_{\sun}$~yr$^{-1}$~$<$~SFR~$<7.5$~M$_{\sun}$~yr$^{-1}$ and for the most massive galaxy we find a SFR based on the total far-infrared luminosity of $(50 \pm 10)$~M$_{\sun}$~yr$^{-1}$.

The most massive galaxy is also detected in our ALMA $^{12}$CO(2--1) and $^{12}$CO(3--2) observations, from which we derive a molecular gas mass of M$_{\rm{H2}} = 1 - 9 \times 10^{10}$~M$_{\sun}$. We infer that this galaxy is likely to be a LIRG based on the emission line ratios as well as the ratio of the far-infrared based SFR and the dust-uncorrected SFR based on the $[$OII$]$ emission line flux. This is the first time that multiple $^{12}$CO transitions are observed from one galaxy connected to a Ly~$\alpha$ absorber. For the other galaxies in the field we determine an upper limit of M$_{\rm{H2}} < 17 \times 10^{9}$~M$_{\sun}$.

We are able to model the morpho-kinematics of the three closest galaxies. As presented in \ref{SecGalfit}, all three galaxies have velocity fields consistent with a rotating disk. Moreover, we construct a line of sight velocity map from the $^{12}$CO(2--1) emission line for \GalB. It is found that the velocity field from the cool ionized gas and the molecular gas are comparable.

We explore different explanations for the neutral gas probed in absorption.
\begin{itemize}
\item Another galaxy could be closer to the line of sight towards the quasar, that is not seen in our observations. The limiting dust-uncorrected SFR in our MUSE data cubes is $0.2 \text{ M}_{\sun} \text{ yr}^{-1}$ in the field and $0.3 \text{ M}_{\sun} \text{ yr}^{-1}$ at the position of the quasar. At this low SFR, the galaxy is unlikely to have an outflow and therefore it must be close to the quasar line of sight and the absorption is actually probing the ISM of this galaxy. We note that we cannot rule out this possibility completely, but we emphasize that our MUSE observations reach a higher completeness compared to previous studies using broad-band imaging.
%\item Comparing the velocity at which we see the absorber with the extrapolation of the rotating disk model suggests that the intervening absorber could trace a very extended disk of \GalB, for which the impact parameter is $133$~kpc. However, such a large disk is unrealistic and we do not believe this explanation.
%\item The velocity at which we see the absorber is inconsistent with being part of a very extended disk of the two closest galaxies at impact parameters of 102~kpc and 133~kpc.
\item The neutral gas could be tracing an outflow from the most massive galaxy \GalB\ since the quasar line of sight is aligned with the minor axis of the galaxy. We estimate, that for a constant outflow speed of $300$~km~s$^{-1}$ it would take the gas $400$~Myr to travel to the position of the absorber. However, the question remains how the gas can stay cool in the outflow or whether it can cool at larger distances. Furthermore, outflows of neutral gas are not yet observed at such large distances.
\item The impact parameter of $133$~kpc for the most massive galaxy is comparable with the spatial extent of the intra-group gas quoted by \mbox{\citet{Bordoloi2011radial}}. Moreover, recent studies frequently find intervening absorbers probing the intra-group gas. We argue that this is, apart from an undetected galaxy closer to the quasar-line-of-sight, the most probable explanation for the gas traced by the absorption.
\end{itemize}

Finally, finding multiple galaxies at the absorber redshift adds further evidence to the findings by \citet{Whiting2006mos, Kacprzak2010galaxy, Gauthier2013ultra, Rahmani2017observational, Bielby2017probing, Fumagalli2017witnessing, Peroux2017nature} suggesting that the classical picture of a one to one correlation between a Ly~$\alpha$ absorber and the host galaxy is incomplete and needs to be revised.

%- Based on the metallcity extrapolated to the absorber prosition it is unlikely that it is an extendet disk from any of the galaxies (preliminarly, comparing different pahses of the gas)

%\begin{itemize}
%\item this is a case study showing the power of combining MUSE and ALMA observations of the $^{12}$CO emission lines of galaxies connected with a Ly~$\alpha$ absorber.
%\item Molecular gas: 
%\begin{itemize}
%\item probe the molecular gas content of galaxies identified as a Ly~$\alpha$ absorber
%\item when observing multiple $^{12}$CO emission lines it is possible to constrain the ISM %properties from the $^{12}$CO SLED
%\end{itemize}
%\item HII regions and neutral gas:
%\begin{itemize}
%\item identify all galaxies at the absorber redshift efficiently
%\item probe the ionized gas
%\item study the morpho-kinematics
%\end{itemize}
%\item Combination offers a complete baryonic census of the system helping to gain a better understanding of what types of galaxies are probed by a Ly~$\alpha$ absorber
%\end{itemize}

%\section{TODO (Ignore!)}

%\begin{itemize}
%\item put labels of galaxies in Fig. 11
%\item put labels of lines in Fig. 10
%\item submit 43 to GAZPAR - DONE waiting for the results
%\item tabulate the absorber fit properties
%\item show kinematics on the MUSE image not so easy...
%\item present this as a case study of combining ALMACAL and MUSE
%\item get metallicity for 12 or all galaxies from mass metallicity relation - Zahid et al. 2011 use le phare to get stellar masses!
%\end{itemize}

\section*{Acknowledgements}

We thank the anonymous referee for a constructive report. The authors thank Chris Churchill for kindly providing us with the reduced HIRES spectrum of the quasar and Sandhya Rao for kindly providing us with the broad band images. We also thank Nicolas Bouch\'e for his advice on the morpho-kinematic modelling using Galpak$^{\rm 3D}$. The authors thank Hadi Rahmani for useful discussion about intra-group medium. We are grateful to the ESO staff at Paranal and in Garching for performing the MUSE observations in service mode. 
AK acknowledges support from the STFC grant ST/P000541/1 and Durham University. 
CP was supported by an ESO science visitor programme and the DFG cluster of excellence `Origin and Structure of the Universe'. 
IRS acknowledges support from the ERC Advanced Grant DUSTYGAL (321334), a Royal Society/Wolfson Merit Award and STFC (ST/P000541/1).
IO acknowledges support from the European Research Council in the form of the Advanced Investigator Programme, 321302, {\sc cosmicism}.
This paper makes use of the following ALMA data: \\
ADS/JAO.ALMA\#2015.1.00920.S, 
ADS/JAO.ALMA\#2015.1.00262.S, 
ADS/JAO.ALMA\#2012.1.00350.S, 
ADS/JAO.ALMA\#2013.1.00403.S. 
ADS/JAO.ALMA\#2013.1.00111.S, 
ADS/JAO.ALMA\#2013.1.01175.S,
ADS/JAO.ALMA\#2013.1.01175.S,
ADS/JAO.ALMA\#2013.1.01225.S, 
ADS/JAO.ALMA\#2012.1.00146.S, 
ADS/JAO.ALMA\#2015.1.00350.S, 
ADS/JAO.ALMA\#2013.1.00198.S, 
ADS/JAO.ALMA\#2013.1.01172.S, 
ADS/JAO.ALMA\#2013.1.00815.S, 
ADS/JAO.ALMA\#2013.1.00815.S, 
ADS/JAO.ALMA\#2013.1.00710.S, 
ADS/JAO.ALMA\#2015.1.00920.S, 
ADS/JAO.ALMA\#2016.1.00724.S, 
ADS/JAO.ALMA\#2016.1.00683.S, 
ADS/JAO.ALMA\#2016.1.00627.S, 
ADS/JAO.ALMA\#2016.1.00627.S, 
ADS/JAO.ALMA\#2016.1.01453.S, 
ADS/JAO.ALMA\#2016.1.01262.S, 
ADS/JAO.ALMA\#2015.1.00296.S,
ADS/JAO.ALMA\#2013.1.01175.S,\\
and ADS/JAO.ALMA\#2012.1.00650.S.

ALMA is a partnership of ESO (representing its member states), NSF (USA) and NINS (Japan), together with NRC (Canada), MOST and ASIAA (Taiwan), and KASI (Republic of Korea), in cooperation with the Republic of Chile. The Joint ALMA Observatory is operated by ESO, AUI/NRAO and NAOJ. This work makes use of observations made with the NASA/ESA Hubble Space Telescope, or obtained from the data archive at the Space Telescope Institute (STScI), which is a collaboration between STScI/NASA, the Space Telescope European Coordinating Facility (STECF/ESA) and the Canadian Astronomy Data Centre (CADC/NRC/CSA). This research made use of Astropy, a community-developed core Python package for Astronomy \citep{Astropy2013astropy}.

%The Acknowledgements section is not numbered. Here you can thank helpful
%colleagues, acknowledge funding agencies, telescopes and facilities used etc.

%%%%%%%%%%%%%%%%%%%%%%%%%%%%%%%%%%%%%%%%%%%%%%%%%%



%\bibliographystyle{mnras}
%\bibliography{Bibliography} % if your bibtex file is called example.bib

%\begin{thebibliography}{99}
%\end{thebibliography}

%%%%%%%%%%%%%%%%%%%%%%%%%%%%%%%%%%%%%%%%%%%%%%%%%%

%%%%%%%%%%%%%%%%% APPENDICES %%%%%%%%%%%%%%%%%%%%%

\appendix



\section{Morphology and Kinematics}

\begin{figure}
\includegraphics[width = \linewidth, trim=0 400 0 0 crop]{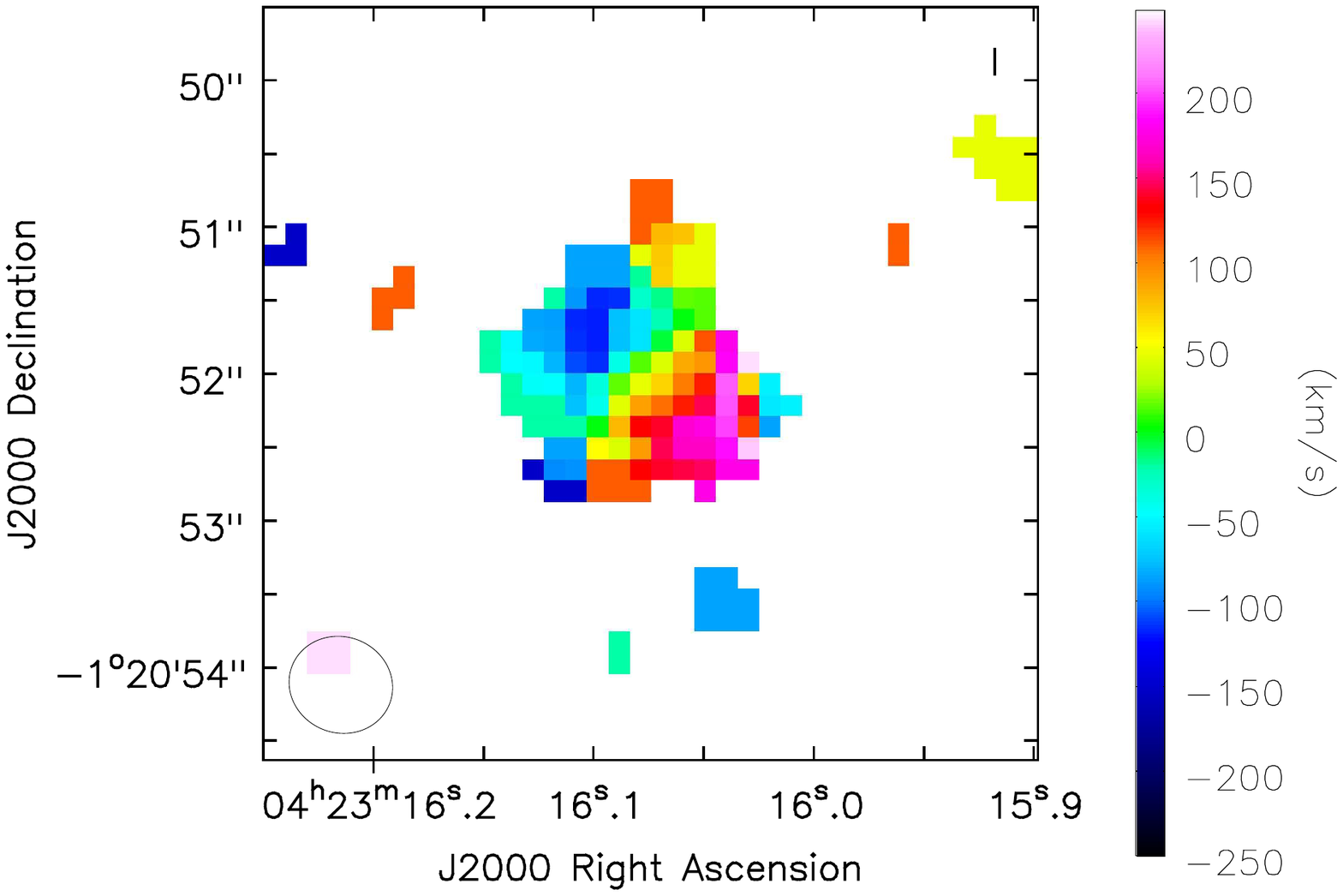}
\caption{The intensity weighted line-of-sight velocity field of the $^{12}$CO(2--1) emission from the ALMA observations for \GalB. The velocity map is consistent with a rotating disk and the orientation is consistent with the model velocity field based on [OIII] emission shown in Fig.~\ref{FigMUSEvelFieldOverview}.}
\label{FigCO2--1VelMap}
\end{figure}

\begin{figure*}
\centering
\begin{minipage}{0.40\linewidth}
\centering
\begin{tikzpicture}
    \node[anchor=south west,inner sep=0] (image) at (0,0) {\includegraphics[width = \linewidth, trim=0 35 0 35,clip]{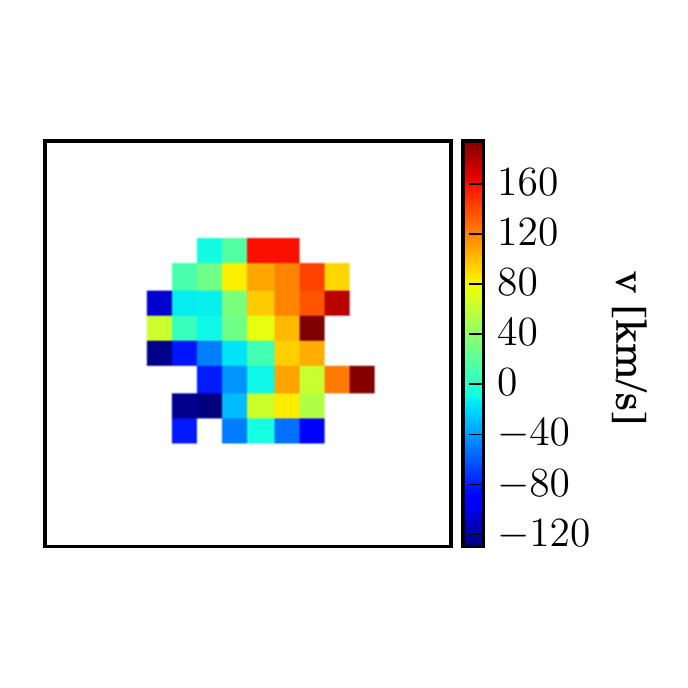}};
    %\node [anchor=west] (QSO) at (3.4,3.8) {\Large backgr.\\QSO};
    \begin{scope}[x={(image.south east)},y={(image.north west)}]
        \draw [-latex,ultra thick] (0.4,0.7) -- node[right]{\Large QSO} (0.45,0.9);
        \node[align=left] at (0.165, 0.8) {\Large J0423A};
    \end{scope}
\end{tikzpicture}
\end{minipage}
\begin{minipage}{0.25\linewidth}
\centering
\includegraphics[width = \linewidth]{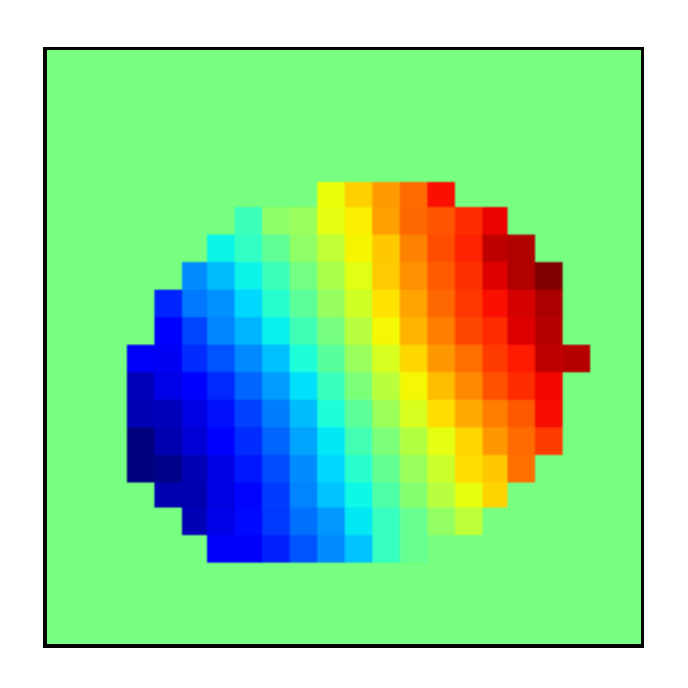}
\end{minipage}
\begin{minipage}{0.25\linewidth}
\centering
\includegraphics[width = \linewidth]{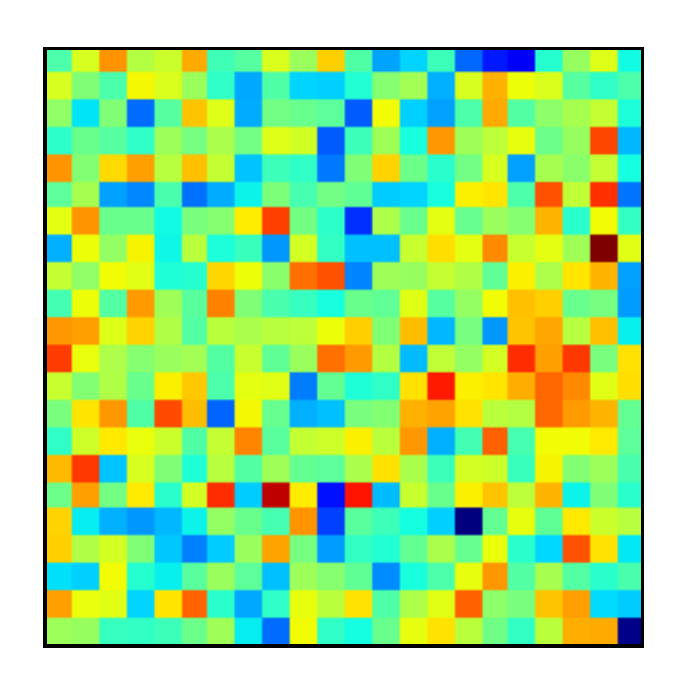}
\end{minipage}\\
\begin{minipage}{0.40\linewidth}
\centering
\begin{tikzpicture}
    \node[anchor=south west,inner sep=0] (image) at (0,0) {\includegraphics[width = \linewidth, trim=0 35 0 35,clip]{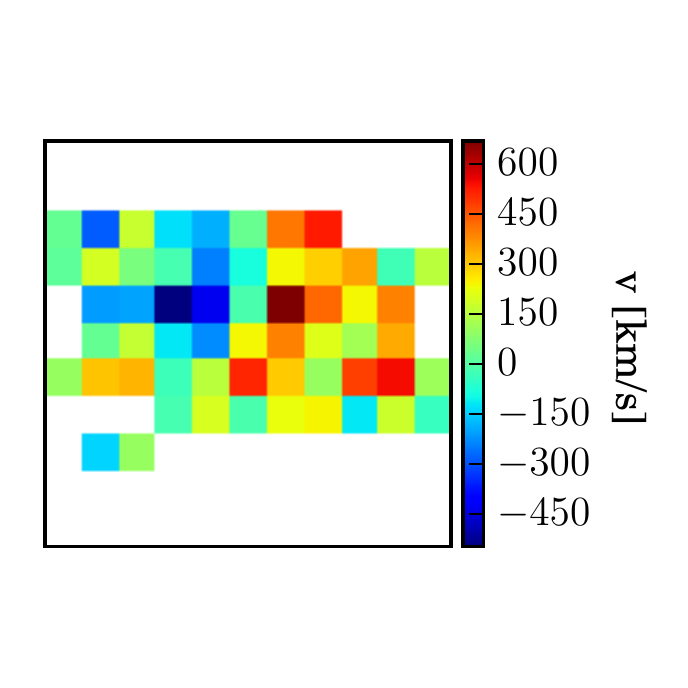}};
    %\node [anchor=west] (QSO) at (3.4,3.8) {\Large QSO};
    \begin{scope}[x={(image.south east)},y={(image.north west)}]
        \draw [-latex,ultra thick] (0.4,0.7) -- node[right]{\Large QSO} (0.425,0.9);
        \node[align=left] at (0.165, 0.8) {\Large J0423B};
    \end{scope}
\end{tikzpicture}
\end{minipage}
\begin{minipage}{0.25\linewidth}
\centering
\includegraphics[width = \linewidth]{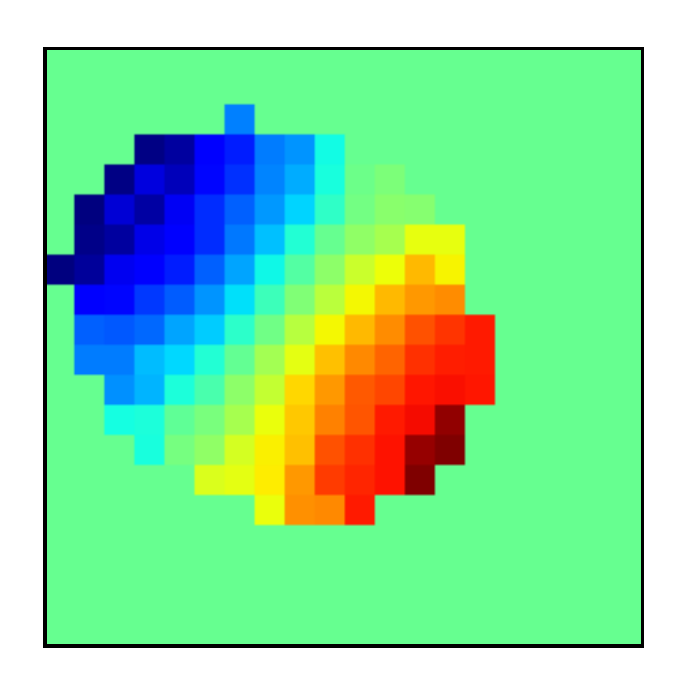}
\end{minipage}
\begin{minipage}{0.25\linewidth}
\centering
\includegraphics[width = \linewidth]{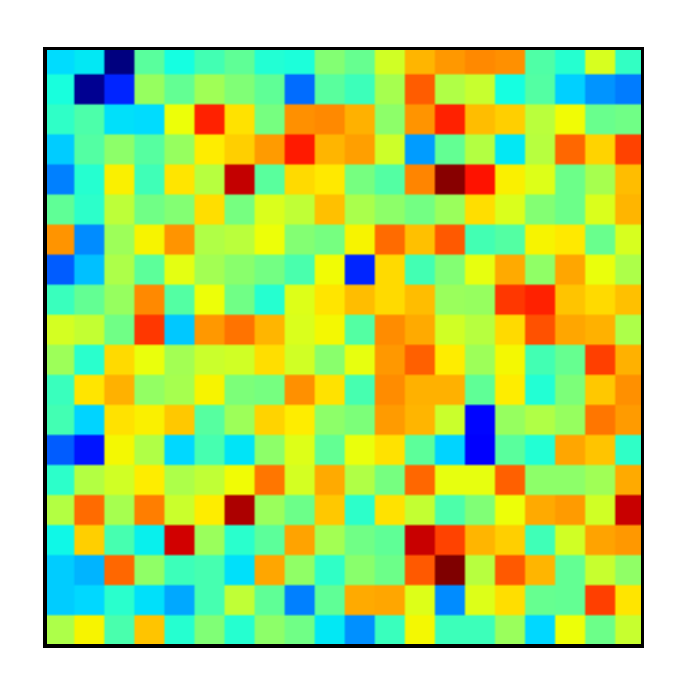}
\end{minipage}\\
\begin{minipage}{0.40\linewidth}
\centering
\begin{tikzpicture}
    \node[anchor=south west,inner sep=0] (image) at (0,0) {\includegraphics[width = \linewidth, trim=0 35 0 35,clip]{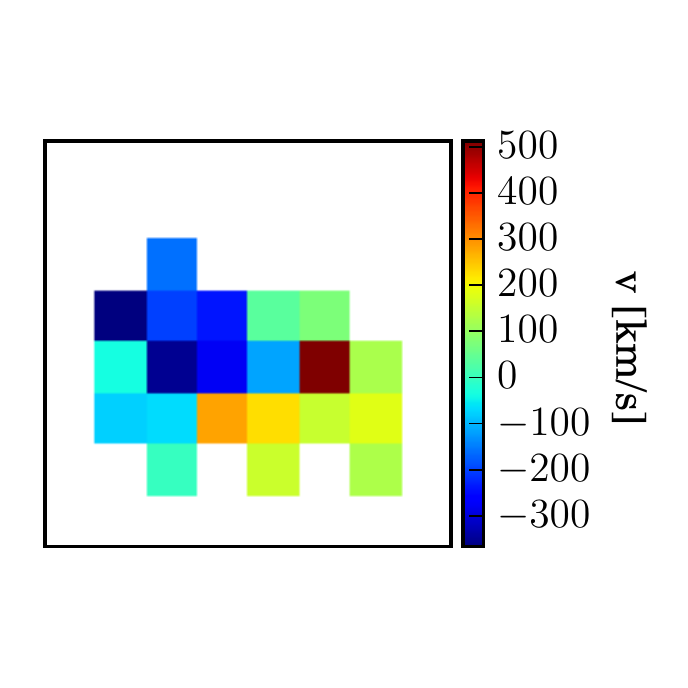}};
    %\node [anchor=west] (QSO) at (3.4,3.8) {\Large QSO};
    \begin{scope}[x={(image.south east)},y={(image.north west)}]
        \draw [-latex,ultra thick] (0.3,0.7) -- node[right]{\Large QSO} (0.26,0.9);
        \node[align=left] at (0.165, 0.8) {\Large J0423C};
    \end{scope}
\end{tikzpicture}
\end{minipage}
\begin{minipage}{0.25\linewidth}
\centering
\includegraphics[width = \linewidth]{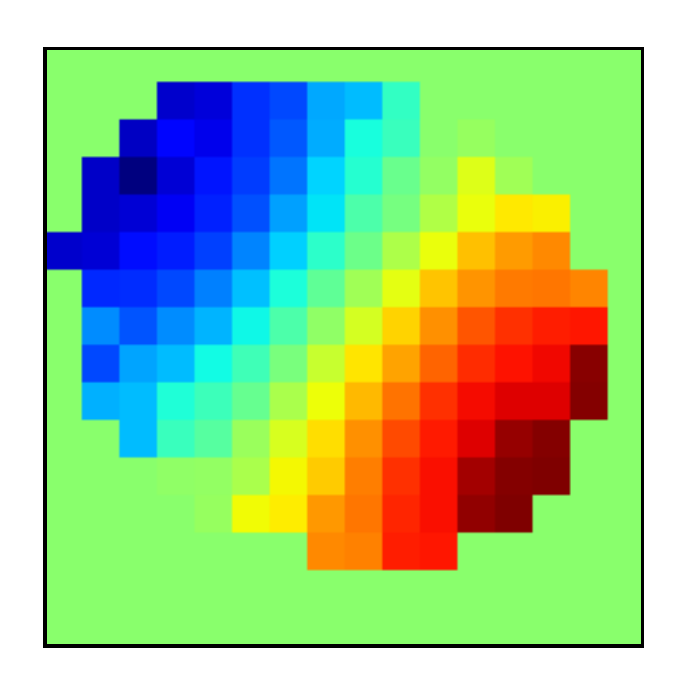}
\end{minipage}
\begin{minipage}{0.25\linewidth}
\centering
\includegraphics[width = \linewidth]{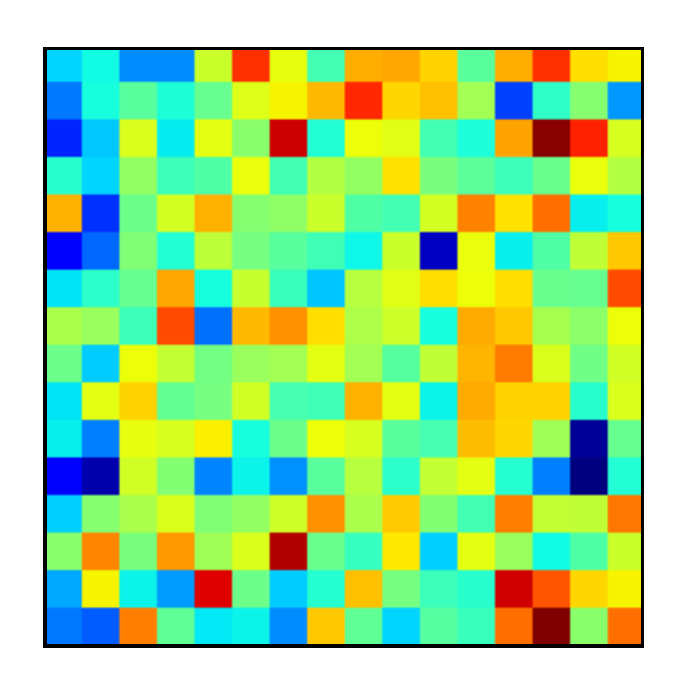}
\end{minipage}\\
\caption{Results from the morpho-kinematic fitting to the [OIII] emission detected with MUSE, including the observed velocity field~(left), the model velocity field from {\sc GalpaK$^{\rm 3D}$} fitting~(middle) and the residual map (right) for \GalA\ (top), \GalB\ (middle) and \GalC\ (bottom). The arrow indicated the direction towards the quasar sight-line. We find that all three galaxies have velocity fields consistent with rotating disks. We caution that the maximum velocity from the modelling is uncertain because all three galaxies are very compact ($<3$~kpc) and {\sc GalpaK$^{\rm 3D}$} is known to overestimate the maximum velocity in the case of compact galaxies \citep{Bouche2015galpak}. We also note that the observed velocity fields are spatially re-sampled for display purpose.}
\label{FigMUSEvelFieldOverview}
\end{figure*}

To study the kinematics of the molecular gas, we create a line-of-sight velocity map using the CASA task {\sc IMMOMENTS} including pixels where the $^{12}$CO line emission is detected at $3\sigma$ above the noise in each velocity channel of the cube. The resulting map is shown in Fig.~\ref{FigCO2--1VelMap}. The velocity field is sampled with only three resolution elements across the major axis, but given this limitation, the velocity field is consistent with that of a rotating disk.\\

\label{SecGalfit}

To determine the morphology and kinematics of the four $[$OIII$]$-detected galaxies in our MUSE cube, we perform a two-step analysis. First, we perform a Sersic-fit of the [OIII] pdfeudo-narrow-band image using {\sc Galfit} \mbox{\citep{Peng2002detailed}}. Second, we use the {\sc GALPAK$^{\rm 3D}$} algorithm \mbox{\citep{Bouche2015galpak}} to perform a 3D morpho-kinematic fit to the [OIII] emission line cube. We use the Sersic profile as input for the 3D morpho-kinematic fit. We infer from this fit whether the kinematics are compatible with that observed in a rotating disk and how the position angles of the galaxies are related to each other and how the quasar absorption is oriented with respect to the major and minor axes of the galaxies. However, given the data quality we will not be able to detect warpdf in the disks.

We determine the half-light radius, axis ratio, position angle (PA) and the Sersic index from the [OIII] pdfeudo-narrow-band image using {\sc Galfit}. This is a two-dimensional fitting algorithm extracting structural parameters from galaxy images. The algorithm models the light profile and is designed to fit multiple components. Here we perform a single component fit of the [OIII] pdfeudo-narrow band image. The fitting results are shown in Table \mbox{\ref{TabParamGalfit}}. We find that for \GalD\ the PA and Sersic index are not very well constrained due to the low SNR of the detected line emission.

\begin{table*}
\begin{minipage}{\linewidth}
\centering
\caption{Summary of the morphological parameters from 2D fitting and morpho-kinematic parameters from 3D fitting.}
\label{TabParamGalfit}
\label{TabKinematicParam}
\begin{tabular}{l c c c c c c c c c}
\hline
Name & $r_{1/2}^{\rm Galfit}$ & $r_{1/2}^{\rm GalpaK}$ & $PA^{\rm Galfit}$ & $PA^{\rm GalpaK}$ & $i^{\rm Galfit}$ & $i^{\rm GalpaK}$ & Sersic index & $v_{\rm max}^{\rm GalpaK}$ & $\sigma_0^{\rm GalpaK}$\\
&[kpc] & [kpc] & [$^{\circ}$] & [$^{\circ}$] & [$^{\circ}$] & [$^{\circ}$] & & [km s$^{-1}$] & [km s$^{-1}$]\\
\hline
\GalA\ & $1.5 \pm 0.6$ & $2.9 \pm 0.3$ & $106 \pm 4$ & $115 \pm 4$ & $33 \pm 3$ &  $42 \pm 6$ & $3 \pm 1$ &$< 410$ & $11 \pm 7$ \\
\GalB\ & $2.77 \pm 0.08$ & $2.77^{\star}$ & $90 \pm 2$ & $54 \pm 24$ & $ 40.5 \pm 0.8 $ & $51 \pm 17$ & $1.04 \pm 0.03$ & $<170$ & $84 \pm 17$\\
\GalC\ & $1.54 \pm 0.09$ & $2.73 \pm 0.03$ & $\: 67 \pm 6$ & $54 \pm 21$ & $47 \pm 3$ & $50 \pm 16$ & $0.71 \pm 0.06$ & $<191$ & $87 \pm 17$ \\
\GalD\ & $0.5 \pm 0.1$ & & $-90 \pm 90$ & & $60 \pm 30$ & & $0.3 \pm 0.4$\\
\hline\\
\end{tabular}
\end{minipage}
\begin{minipage}{\linewidth}
Note: $^{\star}$ is fixed based on the 2D fitting from {\sc Galfit}. We find that for \GalA, \GalB\ and \GalC\ the morphological parameters from independent 2D and 3D fitting are comparable.
\end{minipage}
\end{table*}

%\subsubsection{Emission Line Width and Ionization Properties}
\label{SecLineWidthMEx}

The GALPAK$^{\rm 3D}$ algorithm is used to derive the kinematics of the galaxies based on the [OIII] emission line. We do not use the [OII] emission line, because it is an unresolved doublet and we do not know the exact line ratio.

The GALPAK$^{\rm 3D}$ algorithm directly compares a number of parametric models, created from a Markov Chain Monte Carlo (MCMC) algorithm and mapped in $x, y, \lambda$ coordinate system, to the data. 
GALPAK$^{\rm 3D}$ fits the model in three dimensions and offers even in poor seeing conditions a robust determination of the morpho-kinematics. It probes the posterior possibility density distribution via an MCMC chain of $15,000$ runs and fits $10$ parameters simultaneously (position ($x, y, \lambda$), flux, half-light radius, inclination, PA, turnover radius, maximum velocity and intrinsic velocity dispersion).

The algorithm can only converge if the maximum SNR $> 3$ per pixel, which is not fulfilled for \GalD. Furthermore, the maximum velocity is overestimated if the ratio of galaxy half-light radius to seeing radius is smaller than about $1.5$.

We use the half-light radius from the 2D fitting as a fixed input parameter for \GalB. For \GalA\ it is necessary to fix the turnover radius to break the degeneracy with the maximum velocity. We set the turnover radius to $0.9$ times the half-light radius based on the scaling relation found in local disc galaxies \citep{Amorisco2010self}. Since all of the galaxies are found to be compact already from independent 2D profile fitting, we expect that the GalpaK$^{\rm 3D}$ will overestimate the maximum velocity. The final fitting parameters are given in Table \ref{TabKinematicParam}. We find that the velocity fields of the galaxies \GalA, \GalB, and \GalC\ are consistent with rotating disks. As can be seen in Table \ref{TabKinematicParam} the PAs from different objects are not aligned. Comparing the PA derived with Galpak$^{\rm 3D}$ with the PA from the 2D fitting with {\sc Galfit} we find that they are comparable within the errors. %if we take into account that in the 2D fitting we cannot distinguish between PA and PA~+~180. 
We show the observed and model velocity fields from the MUSE observations in Fig.~\ref{FigMUSEvelFieldOverview}. It can be seen that in \GalB\ the velocity field of the ionized gas traced by the [OIII] emission line and of the molecular gas traced by the $^{12}$CO(2--1) emission line shown in Fig.~\ref{FigCO2--1VelMap} are consistent. Furthermore, we find that the angular extent of $1.4\arcsec \pm 0.1\arcsec$ for the [OIII] emission line matches the angular extent of $1.3\arcsec \pm 0.2\arcsec$ found for the CO(2--1) emission line.\\

We explore the possibility that the absorption in the quasar spectrum is tracing the extended rotating gas disk of one of the two closest galaxies \GalA\ and \GalB. Therefore, we extrapolate the model velocity field derived above to the position of the absorber. The expected velocity is overplotted in the velocity space of the absorber on the MgII$2796$\AA\ absorption line in the last panel of Fig.~\ref{FigAbsLineKineComp}. It can be seen that the expected velocity from \GalA\ does not match the absorption profile. For \GalB\, on the other hand, we find that it agrees very well with one of the absorption components in the MgII$2796$\AA\ absorption profile. Intriguingly, this galaxy is the one for which we measure a large cold gas content. We note, however, that projected separation between the quasar sight line and the gas-rich galaxy is $133$~kpc, so the simple extrapolation of a rotating disk is a very naive assumption.

\section{Neutral Gas Properties}
\label{SecHiresSpec}

In this section, we use the quasar spectrum presented by \mbox{\citet{Churchill1996spatial}} to fit the absorption lines, determine the minimum number of components needed to reproduce the absorption profile and determine their redshift. We will use these fits in Section~\ref{SecHIRESMetallicity} to derive a lower limit on the total metallicity of the absorbing neutral gas. We use the kinematic information and the metallicity of the absorber to infer the alignment with the galaxies in velocity space as well as the possibility of probing an extended galactic disk with the absorption.

\begin{figure*}
\includegraphics[width = \linewidth]{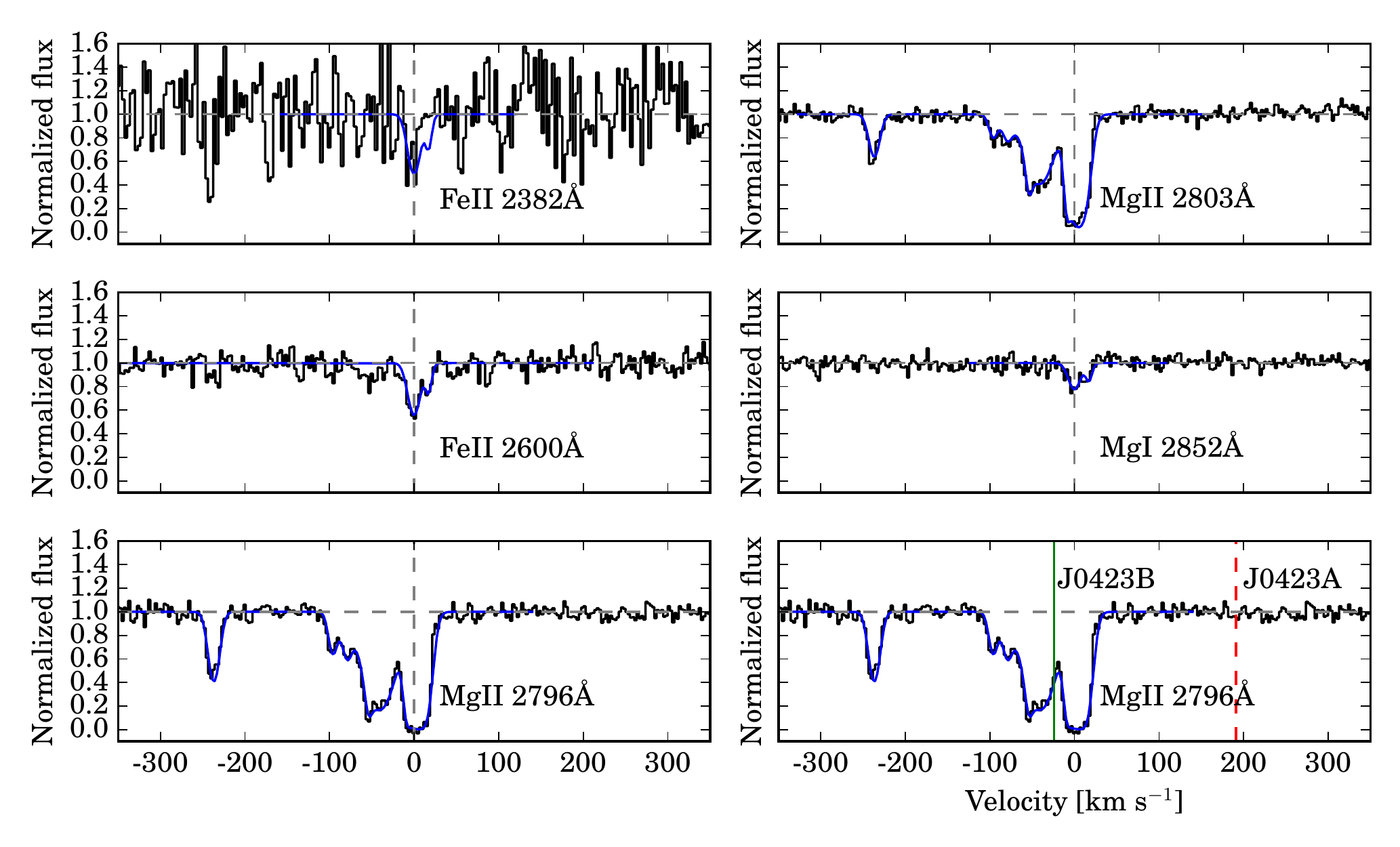}
\caption{Fits to the FeII$2382$\AA, FeII$2600$\AA, MgII$2796$\AA, MgII$2803$\AA, MgI$2852$\AA\ absorption lines in the QSO spectrum. In the fitting using vpfit we assumed two components for FeII and MgI and six components for the saturated MgII. Velocity zero corresponds to the redshift of the main component from the joint fits of FeII$2600$\AA, FeII$2382$\AA, MgI$2852$\AA. The last panel shows a comparison of the MgII$2796$\AA\ absorption line profile with the expected absorption caused by an extended rotating disk of \GalA\ and \GalB\ at the position of the quasar sight-line. We find that the absorption cannot be caused by an extended disk of \GalA, but the velocity extrapolated from the disk of \GalB\ matches the absorption profile. We note, however, that due to the large impact parameter of $133$~kpc the absorption is unlikely tracing the disk of \GalB.}
\label{FigQSOAbsLines}
\label{FigAbsLineKineComp}
\end{figure*}

\begin{table}
\begin{minipage}{\linewidth}
\centering
\caption{Fit parameters for the absorption line Voigt profile fitting using {\sc vpfit}.}
\label{TabAbsFitParam}
\begin{tabular}{l l l l l}
\hline
Ion & No. & $z$ & $b$ & log $N$ \\
& & & & [atoms cm$^{-2}$]\\
\hline
FeII & 1 & $0.633174 \pm 0.000006$ & $3.0 \pm 2.4$ & $12.21 \pm 0.13$\\
FeII & 2 & $0.633080 \pm 0.000005$ & $8.9 \pm 1.4$ & $12.80 \pm 0.05$\\
MgI & 1 & $0.633174 \pm 0.000006$ & $3.0 \pm 2.4$ & $11.04 \pm 0.15$\\
MgI & 2 & $0.633080 \pm 0.000005$ & $8.9 \pm 1.4$ & $11.50 \pm 0.07$\\
\hline
MgII & 1 & $0.631791 \pm 0.000002$ & $7.7 \pm 0.5$ & $12.49 \pm 0.02$\\
MgII & 2 & $0.632558 \pm 0.000005$ & $5.7 \pm 1.3$ & $12.09 \pm 0.06$\\
MgII & 3 & $0.632648 \pm 0.000005$ & $7.0 \pm 1.8$ & $12.16 \pm 0.08$\\
MgII & 4 & $0.632788 \pm 0.000003$ & $2.9 \pm 1.4$ & $12.36 \pm 0.09$\\
MgII & 5 & $0.632855 \pm 0.000004$ & $20.6 \pm 0.9$ & $13.17 \pm 0.02$\\
MgII & 6 & $0.633108 \pm 0.000003$ & $11.5 \pm 0.4$ & $13.49 \pm 0.04$\\ 
\hline
\end{tabular}
\end{minipage}
\begin{minipage}{\linewidth}
Note: The top four rows above the dividing line show the parameters for the Voigt Profile fitting using only FeII and MgI. The bottom six lines below the dividing line show the fit parameters for the Voigt profile fit using only the MgII absorption lines. No. denotes the number of the component that is simultaneously fit in multiple absorption lines.
\end{minipage}
\end{table}

We model the metal absorption lines in the quasar spectrum associated with the Lyman Limit system with Voigt profiles using {\sc VPFIT}\footnote{\url{httpdf://www.ast.cam.ac.uk/~rfc/vpfit.html}}v.10.2. The {\sc VPFIT} code is developed to fit multiple Voigt profiles to spectroscopic data by minimizing the $\chi^2$. Here, we have performed a multicomponent fit assuming that for each absorption component all ions have the same redshift. We show the spectra and the Voigt profile fits in Fig. \ref{FigQSOAbsLines}. It can be seen that the MgII$2803$\AA, and MgII$2796$\AA\ absorption lines are saturated and more complex than the FeII$2600$\AA, FeII$2382$\AA, and MgI$2852$\AA\ absorption lines. Therefore, we fit the two sets of absorption lines separately. We find that the absorption lines are well fitted using two-components for the FeII and MgI lines. A separate fit is performed for the strong MgII lines using six components. We summarize the corresponding fit parameters in Table \ref{TabAbsFitParam}.

%\section{Analysis}

\section{Metallicities}

We study the metallicity of the ionised gas in the galaxies and of the neutral gas traced by the absorption. Using this information, we can compare the metallicities observed at different positions with each other and infer from this the connection between the galaxies and the absorbing gas.

\subsection{HII Metallicities}
\label{SecEmLinesMetallicity}

To determine the gas metallicities of the MUSE-detected galaxies from the emission lines, we use the R23 method as first introduced by \citet{Pagel1979composition}, which is widely used to determine gas metallicities if the H~$\beta$, [OII], and [OIII] fluxes are known. It uses the $\log R_{23}$ parameter, which is defined as follows:

\begin{equation}
\log(R_{23}) = \log \left(\frac{F([OII]\lambda 3727, 3729) + F([OIII]\lambda4959, 5007)}{F(H\beta)}\right).
\end{equation}

We derive the metallicities according to the formalism described as the ``best'' oxygen abundance determined by \mbox{\citet{Kobulnick2004metallicities}}. 
We use the O32 parameter ($f_{\rm [OIII]}/f_{\rm[OII]}$) as a branch indicator \mbox{\citep{Maiolino2008amaze}}, where O32~$>2$ suggests the lower branch solution and O32~$<1$ suggests the upper branch solution. The four galaxies in this study have O32 values between $0.26$ and $0.74$ indicating that the upper branch solution is appropriate. The derived metallicities of our galaxies are listed in \mbox{Table~\ref{TabEmissionLines}}. 
Compared to the solar abundance of \mbox{log(O/H) + $12 = 8.69$} \mbox{\citep{Asplund2009chemical}} we find that galaxy \GalA\ and \GalC\ have metallicities of $0.11$ and $0.25$. However, since we are not applying a dust correction we could overestimate the metallicities.

%\subsubsection{Stellar mass based metallicity}
\label{SecMassMetallicity}

Since we do not detect the H~$\beta$ emission line from \GalB\ and \GalD, we use the stellar mass-metallicity relation to determine the gas phase metallicity of these two galaxies. We use the relation given by \mbox{\citet{Zahid2014universal}} who fits the following function to a sample from DEEP2 at $z\simeq 0.8$:

\begin{equation}
12 + \log (\rm{O}/\rm{H}) = Z_0 + \log \left(1 - \exp\left( - \left[\frac{M_{\star}}{M_{0}} \right]^{\gamma}\right)\right) , 
\end{equation}

where $Z_0 = 9.10$,  $\log(M_0 [\text{M}_{\sun}]) = 9.80$ and $\gamma = 0.52$. This yields a metallicity of $12 + \log(\rm{O}/\rm{H}) =  9.1 \pm 0.9$ and $9.0 \pm 0.9$ for \GalB\ and \GalD, respectively. The $1\sigma$ scatter in the DEEP2 data is quite large and therefore the derived metallicity has a large error bar.

%Error in in magnitude from std of background in 3 sigma clipped image 2.5/np.log(10) * (std(flux)*detectArea)/FLUX_AUTO

\subsection{Neutral Gas Metallicity}
\label{SecHIRESMetallicity}

We use the combined fits of FeII$2600$\AA, and FeII$2382$\AA\ and the HI column density derived by \mbox{\citep{Rao2006damped}} to determine the metallicity of the absorbing gas without taking dust into account. We report a lower limit to reflect possible ionisation correction. We find that $[\rm{Fe/H}] = \log(\rm {Fe}/\rm {H})_{\rm abs} - \log(\rm{ Fe}/\rm {H})_{\sun}$ is higher than $-1.16$.

\subsection{Comparison}

\begin{table}
\caption{Metallicity of the galaxies and expected metallicity based on extrapolation using a constant slope at the absorber position assuming an extended gas disk}
\label{TabMetalExtrapolation}
\centering
\begin{tabular}{l l l l l}
\hline
Name & $Z_{\rm mean}$ & $Z_{\rm central}$& $\theta$[kpc] & $Z_{\rm exp}/Z_{\sun}$\\
\hline
\GalA\ & $0.1 \pm 0.1$ & $0.2 \pm 0.1$ & $102$ & $-2 \pm 1$\\
\GalB\ & $0.4 \pm 0.9$ & $0.5 \pm 0.9$ & $133$ & $-3 \pm 1$\\
\GalC\ & $0.25 \pm 0.06$ & $0.32 \pm 0.06$ & $146$ & $-3 \pm 2$\\
\GalD\ & $0.3 \pm 0.9$ & $0.3 \pm 0.9$ & $216$ & $-5 \pm 2$\\
\hline
\end{tabular}
\begin{minipage}{\linewidth}
Note: $Z = 0$ is equal to the solar abundance.
\end{minipage}
\end{table}

Large integral field surveys have shown that galaxies with $\log(M_{\star} )> 9.6$ have a uniform metallicity gradient within the disk of $-0.026$~dex~kpc$^{-1}$ \mbox{\citet{Ho2015Metallicity}}. However, at distances beyond $2R_{\rm e}$ first pieces of evidence for a flattening of the metallicity gradients are found \citep{Belfiore2017sdss}. For absorption line systems a similar shallow negative metallicity gradient is observed out to $25$~kpc \citep[e.g.][]{Peroux2014sinfoni, Christensen2014verifying, Rahmani2017observational}. In the system presented in this work the impact parameter is $>100$~kpc and therefore much larger than the ones probed in the aforementioned studies. We test whether we can connect the galaxies with the absorbing gas using a uniform metallicity gradient.
We extrapolate the metallicity observed in the four galaxies to the position of the intervening absorber using a constant gradient from the literature. First, we derive the central metallicity of the galaxies from the total metallicity reported in Table~\ref{TabPhysParam} and using the solar abundance of \mbox{log(O/H) + $12 = 8.69$} \mbox{\citep{Asplund2009chemical}}. To derive the central metallicity we assume a linear distribution of the mean metallicity quoted in Table~\ref{TabMetalExtrapolation} over two $r_{1/2}$ where we use $r_{1/2}$ from the 3D fitting if available. We use the standard metallicity gradient determined within the disk quoted above. The same metallicity gradient is then used to extrapolate the derived central gas phase metallicity to the position of the absorber. The mean and central metallicity of the galaxies and the expected metallicity at the position of the absorber are given in Table~\ref{TabMetalExtrapolation}. We find that the extrapolated metallicity is above a simple extrapolation of the metallicity gradient. This shows that the metallicity gradient cannot be extended to such large distances.\\

However, here we are comparing the metallicity of the ionized gas in the galaxy with the metallicity of neutral gas probed by the absorption. Furthermore, the standard deviation in the observed gradients is $\pm0.010$~dex~kpc$^{-1}$ in the study by \mbox{\citet{Ho2015Metallicity}}. The corresponding uncertainty in the extrapolation is shown in Table~\ref{TabMetalExtrapolation}. Considering this and the additional caveat of using uniform gradients discussed above we regard this calculation only as an indication.

%%%%%%%%%%%%%%%%%%%%%%%%%%%%%%%%%%%%%%%%%%%%%%%%%%

% Don't change these lines
\bsp	% typesetting comment
\label{lastpage}
\end{document}